\documentclass{aa}
\usepackage{natbib,psfig,graphicx,ulem}
\usepackage{soul,color}
\usepackage{longtable,lscape}
\usepackage{amsmath}

\def\mathnew{\mathsurround=0pt}
\def\simov#1#2{\lower .5pt\vbox{\baselineskip0pt \lineskip-.5pt
\ialign{$\mathnew#1\hfil##\hfil$\crcr#2\crcr\sim\crcr}}}

\begin{document}

\title{A comprehensive picture of baryons in groups and clusters of galaxies}

\author{
T.~F.~Lagan\'a \inst{1} 
\and N. Martinet \inst{2}
\and F. Durret \inst{2}
\and G. B. Lima Neto \inst{1}
\and B. Maughan \inst{3}
\and Y.-Y. Zhang \inst{4}
}

\institute{
IAG, USP, R. do Mat\~ao 1226, 05508-090, S\~ao Paulo/SP, Brazil
\and
UPMC-CNRS, UMR7095, Institut d'Astrophysique de Paris, 98bis Bd Arago, 
F-75014, Paris, France 
\and
H. H. Wills Physics Laboratory, University of Bristol, Tyndall Ave, Bristol BS8 1TL, UK
\and
Argelander-Institut f\"ur Astronomie, Universit\"at Bonn, Auf
  dem H\"ugel 71, 53121 Bonn, Germany
}

\date{Accepted . Received ; Draft printed: \today}

\authorrunning{Lagan\'a et al.}

\titlerunning{The baryon fraction in groups and clusters}

\abstract
% context heading (optional)
%{The baryon mass fraction in groups and clusters is important to set
%  cosmological constraints and should closely match the WMAP-7 value
%  in massive clusters. However, it has been found to decrease towards
%  lower mass systems, with opposite variations of the stellar and
 % X-ray gas mass fractions with total system mass.}
% aims heading (mandatory)
{} {Based on XMM-Newton, Chandra, and SDSS data, we investigate the
  baryon distribution in groups and clusters and its use as a
  cosmological constraint.  For this, we considered a sample of 123
  systems with temperatures  ${\rm kT_{500}} = 1.0-9.0$~keV, total
  masses in the mass range  ${\rm M_{500} = (\sim 10^{13} - 4 \times
  10^{15})\ h_{70}^{-1} \ M_{\odot}}$, and redshifts 0.02 $< z <$1.3.}
% methods heading (mandatory)
{The gas masses and total masses are derived from X-ray data under the
  assumption of hydrostatic equilibrium and spherical symmetry.  The
  stellar masses are based on SDSS-DR8 optical photometric data. For
  the 37 systems out of 123 that had both optical and X-ray data
  available, we investigated the gas, stellar, and total baryon mass
  fractions inside $r_{2500}$ and $r_{500}$ and the differential gas mass
  fraction within the spherical annulus between $r_{2500}$ and $r_{500}$, as a
  function of total mass. For the other objects, we investigated the
  gas mass fraction only.}
% results heading (mandatory)
{We find that the gas mass fraction inside $r_{2500}$ and $r_{500}$
  depends on the total mass.
  %and the observed trend describes
  %well both low and high mass systems, although groups are not the
  %scaled down version of clusters.  
  However, the differential gas mass fraction does not show any
  dependence on total mass for systems with ${\rm M_{500} > 10^{14}
    M_{\odot}}$.  The stellar mass fraction inside $r_{2500}$ and
  $r_{500}$ increases towards low-mass systems more steeply than the
  $f_{\rm gas}$ decrease with total mass.  
 Adding the gas and stellar mass fractions to obtain the total
baryonic content, we find it to increase with cluster mass, reaching
the WMAP-7 value for clusters
  with ${\rm M_{500} \sim 10^{14} M_{\odot}}$.  This led us to investigate
  the contribution of the intracluster light to the total baryon
  budget for lower mass systems, but we find that it cannot account for the difference observed.}
% conclusions
{The gas mass fraction dependence on total mass observed for groups
  and clusters could be due to the difficulty of low-mass systems
  to retain gas inside the inner region ($r < r_{2500}$).  Because of 
  their shallower potential well, non-thermal processes are more
  effective in expelling the gas from their central regions
  outwards. Since the differential gas mass fraction is nearly
  constant, it provides better constraints for cosmology.  Moreover, we find
  that the gas mass fraction does not depend on redshift at a
  $2\sigma$ level.  Using our total $f_{\rm b}$ estimates, our
  results imply $\Omega_{\rm m} < 0.55$, and taking the highest
  significant estimates for $f_{\rm b}$, $\Omega_{\rm m} > 0.22$.}


\maketitle

\section{Introduction}

Considering the hierarchical scenario of structure formation, the
groups we observe today are the building blocks of future galaxy
clusters. Although they collapse and merge to form progressively
larger systems, the observations of groups of galaxies show that they
are not the scaled-down version of galaxy clusters
\citep[e.g.,][]{mulchaey00,ponman03,voit05}. Cluster scaling relations
(e.g. the L--T relation) show deviations from self-similar relations
at the low-mass end \citep[e.g.,][ but see also \citet{eck11}]{voit05},
providing evidence of the importance of baryon physics.

Since galaxy groups are cooler with shallower potential wells,
non-gravitational processes (e.g., galactic winds, cooling, active
galactic nucleus feedback, etc.) play a more significant role in less
massive systems.  Also, the matter composition in groups is different
from that in clusters.  Baryons in galaxy groups and clusters can be
divided into two major components, the hot gas between galaxies
and the stars in galaxies. A minor component is the intra-cluster
light.  In clusters, the gas is the baryon-dominant component
(about 6 times more massive than the stellar mass). While in groups, the gas
mass is significantly lower: it is of
the same order as the stellar mass \citep[e.g.,][]{lagana11} and in
some cases, even lower than the stellar component
\citep[e.g.,][]{giodini09}.
The analysis of both components is critical for observational cosmology
when using groups and clusters to constrain the $\Omega_{m}$ parameter
from the observed baryon mass fraction \citep[e.g.,][]{allen02,V09}.

The total baryon mass fraction is defined as the ratio between the
gas+stars and the total mass:
$f_{\rm b} = (M_{\rm gas} + M_{\rm  star})/M_{\rm tot}$.  
In very massive galaxy clusters ($M_{500} \sim 10^{15} M_{\odot}$), the
baryon content is supposed to closely match the \textit{Seven-year Wilkinson
Microwave Anisotropy Probe} 
(WMAP-7) value
\citep[$f_{\rm b}^{\rm WMAP-7} = 0.169 \pm 0.009$,][]{jarosik11}.
However,  it is found that the
total baryon mass fraction (and the gas mass fraction) decreases towards
low-mass systems
\citep[e.g.,][]{lms03,gonzales07,giodini09,sun09,lagana11,zlp11,sun12}, when analyzing a wide range of masses.  
On the other hand, the stellar mass fraction seems to increase from clusters to
groups of galaxies. A straightforward interpretation can be that
the gas mass fraction is directly related to cooling and to the star formation
rate, and thus, a smaller gas mass fraction in groups may be related to an
efficient cooling.  However, the gas mass fraction can also be affected by
AGN feedback, which, as mentioned before, is more significant in groups
than in clusters. As shown in recent numerical simulations performed by
\citet{puchwein10}, the amount of gas removed by AGN heating from the
central regions of clusters and are driven outwards ($r > r_{500}$)
depends on cluster mass and is higher in low-mass systems.

The baryon fraction in clusters and groups is an important
cosmological probe \citep[e.g.,][]{allen04}, and therefore scaling relations, such
as $f_{\rm gas}$--kT or $f_{\rm gas}$--M500, need to be well understood if we want to use
them as a cosmological tool. We will explore here the baryon
distribution in the form of gas and stellar mass fractions in groups
and compare them with the observed baryon fractions observed in
clusters. We will consider in particular the baryon fractions inside the
characteristic radii $r_{2500}$ and $r_{500}$, which are commonly observed with the
present generation of X-ray telescopes.

Previous works have presented an analysis of the gas fraction in
groups, but few have measured the gas properties up to $r_{500}$:
\citet{V06} and \citet{gonzales07} derived the gas mass fraction
within $r_{500}$ for four low-temperature systems. \citet{sun09}
determined gas properties up to $r_{500}$ for 11 out of 43 groups.
In groups of galaxies, the stellar mass is of the order of
the gas mass. To understand the dependence of the total baryon
fraction on total mass, it is important to compute both components,
the stellar and gas mass fractions, in a homogeneous way.  We thus
present here the analysis of nine galaxy groups based on XMM-Newton and
\textit{Sloan Digital Sky Survey} (SDSS DR8) data, for which we could
reliably measure gas properties up to $r_{500}$.
We also included 114 galaxy clusters from \citet[][ hereafter M12]{M12}
to investigate the gas mass component inside $r_{2500}$ and $r_{500}$,
and for 28 of these 114 clusters, we could also estimate the stellar
mass fraction.

The paper is organised as follows.  The sample is described in
Section~\ref{smp}. The data reduction is divided in two sections:
Section~\ref{sec:stellmass} describes the SDSS-DR8 data reduction, the
colour-magnitude diagrams (CMD) constructed to select group galaxies, and the
procedure to compute the stellar masses; Section~\ref{rx}
describes the X-ray analysis, gas estimates, and total mass estimates. In
Section~\ref{res}, we present our results and compare them to previous
results, which are finally summarized in Sect.~\ref{conc}. A $\Lambda$CDM
cosmology with $H_{0} = 70\ \rm km\ s^{-1}\ Mpc^{-1}$ and $\Omega_{M} = 0.3$
is adopted throughout, and all errors are quoted at the 68\% confidence level.

\section{The sample}
\label{smp}

To analyse the baryon mass fraction dependence (gas and stellar mass)
on total cluster mass, it is important to consider a sample that covers a
wide range of mass, and the objects must have optical and X-ray masses
available. 

In the present work, we analyse the sample of 114 clusters from
\citet{M08}, which were updated in {M12}, in the redshift range $0.11 < z <
1.3$ and with temperatures ranging from 2.0 keV up to 8.9 keV. 
28 out of 114 have optical data available.

To complete our analysis, we also included nine low-mass systems,
which were selected as follows.  We first selected all groups with available XMM-Newton and
SDSS-DR8 data, imposing an X-ray flux limit of $L_{X}=1.9
\times 10^{43} \rm erg/s$, which corresponds to $kT \sim 2\ \rm keV$
\citep{eck11}.  We had 19 groups (without considering the Hickson
groups, which are very particular) fulfilling these
conditions. However, nine out of 19 had very shallow X-ray observations
(low exposure times) that did not allow us to constrain the gas
mass. For this reason, we had to exclude those groups. Another group was
on the border of the X-ray detector, so it was excluded too.

We ended with a large sample of 123 systems (0.02$<z<$1.3) with
total masses between $10^{13} M_{\odot}$ and $4 \times 10^{15}
M_{\odot}$. These groups and clusters are listed in
Tab.~\ref{tab:geral}.  We emphasize that we analysed a sample of nine groups and 28 clusters
for the total baryon budget. For the gas mass
fraction analysis, we worked with 9 groups and 114 clusters from
M12. Thus,  the sample is mainly composed by
clusters in the latter part.

\addtocounter{table}{1}

\section{Optical data analysis and stellar mass determination}
\label{sec:stellmass}

In this section we describe the method adopted to compute the
  total stellar mass (in galaxies) for our nine groups of galaxies and 28
  galaxy clusters from M12.  To be homogeneous,the procedure adopted
  in this work was the same for groups and clusters, and we followed
  the steps described in \citet{giodini09} and \citet{bolzonella10}
  that consider statistical membership, background
  correction, and mass completeness as a function of redshift, and a
  geometrical correction.

We used SDSS-DR8 data, from $DERED$ magnitude tables (already
  corrected for internal galaxy extinction) for all sources in the
  $GALAXY$ catalog.  The $GALAXY$ catalogue is essentially complete
  down to 21.3 i-magnitude (SDSS-DR7
  summary\footnote{http://www.sdss.org/dr7/}).

To obtain stellar masses, we first transformed apparent
  magnitudes to absolute magnitudes with the distance modulus, assuming
  K-correction values, K(z), according to the morphological type
  \citep[tables from][]{poggianti97}:
\begin{equation}
M=m-25 - 5 \log(d_{L}/ 1 Mpc) - K (z), 
\end{equation}
where, $d_{L}$ is the luminosity distance.

Absolute magnitudes are converted to luminosities, assuming an
  absolute magnitude of 4.58 for the Sun in the i-band
  \citep{Blanton03}. Luminosities are then converted to masses
  assuming two different mass-to-light ratios \citep[from][]{kauff03}:
  $(M/L_{i})_{\star}=0.74\ M_{\odot}/L_{\odot}$ for late-type and
  $(M/L_{i})_{\star}=1.70\ M_{\odot}/L_{\odot}$ for early-type
  galaxies, as described in \citet{lagana08}.  
  All morphological classification used here is based on the galaxy distribution in a colour-magnitude diagram (CMD), 
  as explained below. Thus, morphological type simply means ``early'' or ``late'' type galaxies.

\subsection{Completeness}

To compare the stellar masses of groups and clusters, we
  defined the completeness in stellar mass ($\mathcal{M}_{\rm lim}^{\rm star}$),
  or galaxy absolute magnitude, as a function of redshift \citep[as
  adopted in ][]{giodini09,bolzonella10,pozzetti10}.  This is the
  lowest mass at which the galaxy stellar mass function can be
  considered as reliable and unaffected by incompleteness.

\begin{figure}[h]
\centering
\includegraphics[angle=90,width=9.cm]{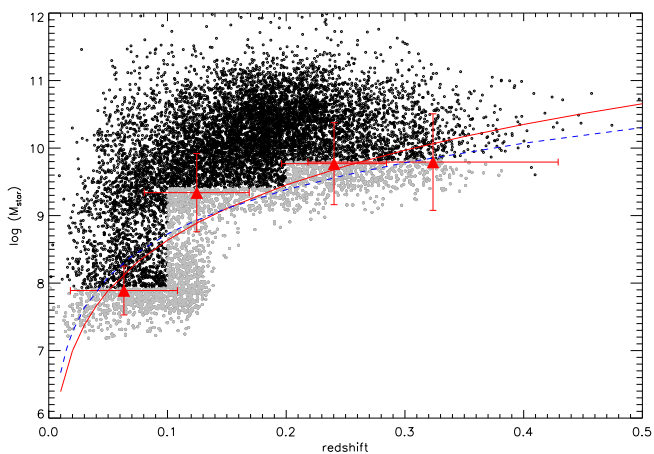}
\caption{Stellar mass completeness for our system, which is computed from the
  fit (red line) to the 95\% percentile of the distribution in the
  limiting mass, for galaxies in the 20\% lower percentile in magnitude
  (grey points) as a function of redshift. The black points represent
  all galaxies inside $r_{\rm 500}$ in the photometric redshift
  range with $i < 21.3$. The blue-dashed line represents the theoretical
  value, assuming no K-correction and a constant value for the mass-to-light ratio, as stated in the text.}
\label{fig:Mlim}
\end{figure}

For each galaxy, we computed the ``limiting mass'', which is
  the stellar mass that this galaxy would have if its apparent
  magnitude was equal to the sample limit magnitude (i.e., i=21.3):
  $\log \mathcal{M}_{\rm lim}^{\rm star} = \log \mathcal{M} +0.4 \times (i-21.3)$, where $\mathcal{M}$
  is the stellar mass of the galaxy with apparent magnitude $i$.
  We then computed this value in small redshift bins by considering
  the 20\%  faintest galaxies (the grey points in
  Fig.~\ref{fig:Mlim}), that is, those contributing to the faint-mass end
  of the galaxy stellar mass function.  For each redshift bin, we
  define the value corresponding to 95\% of the
  distribution of limiting masses as a minimum mass.  The systems were divided into four
  bins of redshift, and we fit the limiting mass values as a function
  of redshift: $ \log \mathcal{M}_{\rm lim}^{\rm star} (z) = 11.67 \times
  z^{0.13}$ (the red line in Fig.~\ref{fig:Mlim}).  We thus adopted
  $\mathcal{M}_{\rm lim}^{\rm star}$ as the lowest galaxy stellar mass that will
  be considered in our analysis to compute the stellar masses for the
  groups and clusters.   As a check, we also computed the theoretical values
  for the limiting mass, assuming
  no K-correction and a constant value for the mass-to-light ratio as 
  $ \log \mathcal{M}_{\rm lim}^{\rm star} \propto DM (z)/2.5$, where DM is the distance modulus.
  We see that the adopted function for the limiting mass (red line in Fig.~\ref{fig:Mlim})  is 
  close to the theoretical predicted value for 
   $\mathcal{M}_{\rm lim}^{\rm star}$ (blue-dashed line in Fig.~\ref{fig:Mlim}).

We estimated the contribution from galaxies that are less massive than $\mathcal{M}_{\rm lim}$ in each redshift bin.
We assumed that the fraction of galaxies that are not are not considered is
$f  = 1 - [\mathcal{M_{\rm star}} (\mathcal{M} > \mathcal{M}_{\rm lim})/\mathcal{M}_{\rm star_{\rm tot}}]$ and can be 
estimated using a stellar mass function. 
Thus,
\begin{equation}
f = 1 - \frac{\int_{\mathcal{M}_{\rm lim}}^{\mathcal{M}_{\rm max}} \mathcal{M} f(\mathcal{M}) d\mathcal{M}}{\int_{\mathcal{M}_{\rm min}}^{\mathcal{M}_{\rm max}} \mathcal{M} f(\mathcal{M}) d\mathcal{M}},
\end{equation}
where $\mathcal{M}_{\rm lim}$ is the limiting mass values for each redshift bin defined by the red points in Fig.~\ref{fig:Mlim},
and $\mathcal{M}_{\rm min}$ and $\mathcal{M}_{\rm max}$ are the minimum and maximum mass values for  the stellar mass in cluster galaxies, respectively.  
We assumed that  $\mathcal{M}_{\rm min} = 10^{8} M_{\odot}$ and $\mathcal{M}_{\rm max} = 10^{13} M_{\odot}$.
Considering that the stellar mass function is described by a single Schechter function, we used typical values from 
\citet{bell03} ($M^{\ast} = 10^{6}/h^{2}$, $\alpha = -0.6$) to estimate the fraction of stellar mass that is missing in each redshift bin because of
completeness correction. We found that the fractional contribution to the total stellar mass budget of galaxies with 
$10^{8} M_{\odot} < \mathcal{M} < \mathcal{M}_{\rm lim} (z)$ varies from less than 1\% up to 4\% (for clusters with $z > 0.3$). 
These small fractions \citep[also in agreement with][]{lms03,giodini09} show that the stellar masses computed in this work are 
reliable.

\subsection{Statistical membership,  background correction and geometrical correction}

As a first step to estimate the projected total stellar mass,
  which is the sum of all potential member galaxies (with masses
  greater than the limiting mass), we first have to define candidate
  members inside $r_{2500}$ and $r_{500}$.  Those members are defined
  as all the galaxies inside a projected distance equal to $r_{2500}$
  or $r_{500}$, which is defined from the X-ray centroid of a group/cluster and with
  photometric redshift in the range $z=z_{\rm c} \pm 0.1 \times (1+
  z_{\rm photo}^{\rm SDSS})$ (where $z_{c}$ is the central redshift taken from NED, and
 $z_{\rm photo}^{\rm SDSS}$ is the photometric redshift taken from SDSS-DR8).  We note
  that we used a range of 0.1, which is of the order of the error on
  photometric redshifts in the SDSS catalogue.

We constructed CMD for each group/cluster to identify the
  red-sequence.  For each system, we linearly fit the red-sequence
  (RS) and considered all galaxies within the RS
  best-fit $\pm 0.3$~mag as early-type objects. Late-type galaxies were classified as the
  objects that are bluer than the lower limit of the RS. We summed the
  masses of all the early-type and late-type galaxies that obeyed our
  criteria.

We corrected for background/foreground
  contamination by measuring the total stellar mass of early-type and
  late-type galaxies, assuming the same photometric redshift
  criterium, in an annulus of inner and outer radii of $8 \times r_{500}$ and
  $9 \times r_{500}$, respectively \citep[as already described
  in][]{lms03,lagana08}.  We applied early- and late-type
  definitions of the cluster to the background region.  Thus, field galaxies are
  selected by following the same criteria as group/cluster potential
  members.  The background regions do not overlap other structures and
  were chosen to represent a field environment.  We then summed the
  masses of all late- and early-type galaxies in the background area.

Finally, we added up the stellar masses of all early-type galaxies
  from our systems and subtracted the stellar masses of background
  early-type galaxies normalised to the cluster area.  The same was done
  for late-type galaxies.  To compute the total stellar mass of
  the system, we summed the total corrected values for early- and late-type 
  stellar masses, considering our limiting mass for the cluster/groups redshift.

There is one last correction to be applied. The values derived
  for the stellar masses refer to a cylinder section projected
  perpendicularly to the line of sight. On the other hand, the gas and
  total masses 
  are measured inside spheres of radii, $r_{2500}$ and $r_{500}$.  To
  compare stellar masses to total and gas masses, we need to apply a
  geometrical correction to correct the cylindrical volume to a spherical
  volume (see Appendix B for more details).  The concentration
  parameters are taken to be 2$\pm$1 and 3$\pm$1 for clusters and
  groups, respectively. These values lead to a multiplicative factor
  on the stellar mass of 0.68 for clusters and 0.74 for groups within
  $r_{500}$ and 0.53 for clusters and 0.61 for groups within
  $r_{2500}$.  One can note that the concentration parameter of
  clusters is chosen to be lower than the parameter of groups, according to
  results from \citet{hansen05}.
  
  \subsection{Uncertainties}
  
  We have three major uncertainties:  the photometric
  magnitude uncertainty (about 14\% of the stellar mass), the
  uncertainty arising from the geometrical correction (about 6\%
  within $r_{500}$ and 9\% within $r_{2500}$), and the uncertainty on
  the mass to light ratio (about 26\%). The error bars are
  approximated to be the quadrature sum of the three uncertainties,
  i.e., about 30\%. The uncertainty on the geometrical correction has
  been calculated, assuming the concentration parameter is known at
  $\pm$1. In most cases, our error bars are mainly because of the
  uncertainty on the mass-to-light ratio. Thus, it could improve with
  better constraints on mass-to-light ratios. It is also interesting
  to note that errors are of the same order within $r_{2500}$ and
  $r_{500}$, which shows there are enough galaxies within $r_{2500}$ to
  calculate statistical uncertainties. Stellar masses with error bars
  are given in Tab.~\ref{tab:geral}.

\section{X-ray data analysis, gas and total mass determinations}
\label{rx}

For the nine groups of our sample, we reduced the X-ray data with the
XMM-Newton Science Analysis System (SAS) v11.0 and calibration
database using all the updates available prior to February 2012.  The
initial data screening was applied using recommended sets of event
patterns - i.e., 0-12 and 0-4 for the MOS and PN cameras, respectively.  The
light curves in the energy range of [1-10] keV were filtered to reject
periods of high background.  We used the background maps for the 3
EPIC instruments from \citet{RP03}. The background was normalised with
a spectrum obtained in an annulus (between 9-11 arcmin), where the
cluster emission is no longer detected. A normalised spectrum was then
subtracted, yielding a residual spectrum. This normalisation parameter
was then used in the spectral fit.
This procedure was already
adopted in \citet{lagana08} and \citet{durret10,durret11}.

We also considered the 114 clusters from \citet{M12}
in the redshift range of $0.1 < z < 1.3$, and observed with Chandra.
Their temperatures ranged from 2.0~keV to 16~keV.  This sample was first 
presented in
\citet{M08}, where the full analysis procedure is described. In M12
the sample was reanalysed with updated versions of the \textit{CIAO}
software package (version 4.2) and the \textit{Chandra} Calibration
Database (version 4.3.0).

\subsection{Gas and total mass determinations for the groups}
\label{secgas}

To compute the gas mass, we first converted the surface
brightness distribution into a projected emissivity profile that was
modelled by a $\beta$-model \citep{CFF78}.  The gas mass is given by:
\begin{equation}
\label{Mgas}
M_{\rm gas}(r) = 4~\pi~m_{p}~\mu_{e} \int_{0}^{r_{\Delta}} n_{e}(r) r^{2} dr, 
\end{equation}
and for the $\beta$-model, we can write
\begin{equation}
n_{e}=\frac{n_{0}}{[1+(\frac{r}{r_c})^2]^{\frac{3\beta}{2}}},
\end{equation}
where $r_{c}$ is the characteristic radius, $\beta$ is the slope
of the surface brightness profile, $\mu_{e}=0.81$, and $n_{0}$ is the central density
obtained from the normalization parameter from the spectra.

To compute the total mass based on X-ray data, we rely on the
assumption of hydrostatic equilibrium (HE) and spherical symmetry. 
The total mass can be calculated using the deprojected surface brightness
and temperature profiles. The total mass is given by:
\begin{equation}
\label{Mtot}
M_{\rm tot} (< r_{\Delta}) = - \frac{k_{b} T r}{G \mu m_{p}}\big(\frac{d \ln \rho}{d \ln r} + \frac{d \ln T}{d \ln r}\big),
\end{equation}
where $r_{\Delta}$ is the radius inside which the mean density is
higher than the critical value by a factor of $\Delta$ (in our case,
$\Delta = 2500$, or $\Delta=500$); $k_{b}$ is the Boltzman constant; T
is the mean gas temperature; $m_{p}$ is the proton mass; $\mu$ is the
molecular weight; and $\rho$ is the gas density. 
Here, we assumed that the systems are isothermal, and we used a global
temperature (measured within $300~h_{70}^{-1} \rm ~kpc$) to compute
the dynamic mass. To determine the temperature, the MOS and PN data were jointly fit 
with a MEKAL plasma model (bremsstrahlung 
plus line emission). 
We fixed the hydrogen column density at the local Galactic
value, using the task \textit{nH} from FTOOLS \citep[an interpolation from the LAB 
\textit{nH} table,][]{kalberla05} to estimate it.
The gas and total masses (derived inside $r_{2500}$
and $r_{500}$) and the other quantities derived from X-ray
observations are in Table~\ref{tab:geral}.

Assuming that the gas temperature for groups is roughly isothermal,
$r_{500}$ is given by \citet{LN03}:
\begin{equation}
\label{detr500}
%r_{\delta}=r_{c} [\frac{4.5 \times 10^{8} \beta <kT>}{\delta h_{70}^{2}E^{2}(z)\mu \r_{c}^2}]^{1/2}
r_{\Delta}=r_{c} \big[\frac{2.3 \times 10^{8} \; \beta <kT>}{\Delta \; h_{70}^{2} E^{2}(z;\Omega_{m}, 
\Omega_{\lambda}) \;  \mu \; r_{c}^{2}}\big]^{1/2},
\end{equation}
where $r_{c}$ is the characteristic radius (given in kpc);  $\beta$ is  
the slope given by the $\beta$-model fit for the surface brightness profile;  $<kT>$ is the mean
temperature (given in keV); and 
$E^{2}(z;\Omega_{m}, \Omega_{\Lambda}) = (\Omega_{m}(1+z)^{3} + (1-\Omega_{m}-\Omega_{\Lambda})(1+z)^{2} + 
\Omega_{\Lambda})$
describes the redshift evolution of the Hubble parameter.
It is important to mention that  the surface brightness and temperature profiles 
reach $r_{500}$ without extrapolation for all systems.

\subsection{Gas and total mass determinations for the \citet{M12} sample}

The gas density profile of each cluster was determined by converting
the observed surface brightness profile (measured in the 0.7-2~keV
band) into a projected emissivity profile, which was then modelled in
M12 by modifying the $\beta$-model \citep[see
e.g.,][]{pointec04,V06,M08} to take into account the
power-law-type cusp instead of a flat core in the centre of relaxed
clusters.
Also, the cluster gas temperature, gas mass and $r_{500}$ were
then determined iteratively in M12.  The procedure was as follows:   to extract a
spectrum within an estimated $r_{500}$ (with the central 15 percent
of that radius excluded), integrate the gas density profile to
determine the gas mass within the estimated $r_{500}$ and thus
calculate $Y_{X}$, which is the product of the temperature and gas mass,
 a low scatter proxy for the total mass \citep{kravtsov06}.  A new
value of $r_{500}$ was then estimated from the $Y_{X} - M_{\rm tot}$
scaling relation of \citet{V09}.  The process was repeated until
$r_{500}$ converged. All the details are in Sect.~2 of M12.

For the
gas mass, we used Eq.~\ref{Mgas}, but we assumed the modified
$\beta$-model (as in M12) given by
\begin{equation}
\label{EqN}
n_{e}^{2}=n_{0}^{2}\frac{(r/r_{c})^{-\alpha}}{(1+r^{2}/r_{c}^{2})^{3\beta-\alpha/2}}\times(1+(r/r_{s})^{\gamma})^{-\epsilon/\gamma},
\end{equation} 
where the additional term describes a change of slope by $\epsilon$
near the radius $r_{s}$, and the parameter $\gamma$ controls the width
of the transition region.
Gas masses were then determined from Monte Carlo realisations of the projected emissivity profile 
based on the best-fitting projected model to the original data. 
The errors in the gas mass determination were calculated using a Gaussian distribution for all gas mass values, and the 
standard deviation was assumed to be the full width at half maximum of this distribution.

\begin{figure}[h]
\centering
\includegraphics[width=9.cm]{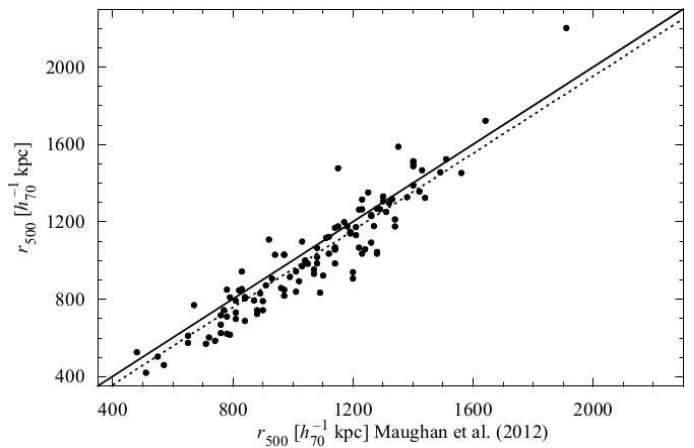}
\caption{Comparison of the $r_{500}$ values determined by \citet{M12}
  and those computed with Eq.~\ref{detr500}  to
  determine $r_{2500}$ and $r_{500}$ in a homogeneous way for our
  sample.}
\label{fig:compr500}
\end{figure}

To test the assumption of isothermality, we computed here the characteristic radius $r_{500}$ and $r_{2500}$,
by its definition,
\begin{equation}
 \langle \rho(r_\delta) \rangle \equiv \frac{M_{\rm tot}(< r_\delta)}{4 \pi r_\delta^3/3} \, ,
\end{equation}
where $\delta$ is the density contrast, $\rho_{\rm cr}(z)$ is the critical density at the cluster redshift, 
and $\langle \rho(r_\delta) \rangle$ is the mean total density inside $r_\delta$. 
The radius $r_\delta$ is then found by numerically solving the following equation:
\begin{equation}
r_\delta^3 = - \frac{3}{4 \pi \delta} \rho_{\rm cr}^{-1}(z) 
\frac{r_\delta \, kT}{\mu m_{\rm H} G} \left.\frac{d \log \rho}{d \log r}\right|_{r=r_\delta} \, .
\end{equation}
Spherical symmetry and hydrostatic equilibrium are assumed, and $kT$ and $\rho$ are the 
gas temperature and density, respectively.
To solve this equation, we used the gas density profile from Eq.~\ref{EqN} and an isothermal temperature.
As shown in Fig.\ref{fig:compr500}, the values
agree with those derived in M12, although  they are systematically lower. They 
may be because the M12 values were derived using a scaling relation,
and we assume hydrostatic equilibrium  (more discussion in Appendix A).

We computed the total mass using Eq.~\ref{Mtot} and assuming 
a modified $\beta$-model for the gas density profile and 
an
isothermal profile, as in the case for the groups. 
In Appendix A,
we test the assumption of isothermality. By showing this, 
we are not introducing any systematic error, and we show that
the total masses derived in this work are robust.
The gas and
total masses inside radii $r_{2500}$ and $r_{500}$ are presented in
Tab.~\ref{tab:geral}. The errors on the total mass determinations 
are mainly because of the assumption of an isothermal gas. 

It is important to state that \citet{piffaretti08} have shown that masses can be 
underestimated by about 5\%, even for the most relaxed clusters. 
\citet{nagai07} found that the gas mass is measured quite accurately ($< \sim 6\%$) 
in all clusters, while the hydrostatic estimate biases the total mass towards lower values by about 
5\%-20\% throughout the virial region, when compared to the gravitational mass estimates.
Thus, we incorporated a 10\% error into our $f_{\rm gas}$ error budget for all 
systems (clusters and groups).

\section{Results}
\label{res}

We present our results in four different sub-sections. In the first
part, we present and discuss the star formation efficiency
 as a function of total mass computed inside $r_{2500}$ and
$r_{500}$ for our sample of 37 systems (nine groups and 28 clusters) with
both optical and X-ray available data. In Sect.~\ref{contr_icl}, we investigate the
difference in mass between the estimated total baryon budget and the
WMAP-7 value. 
Then, we discuss in Sect.~\ref{fracgas} the gas mass fraction enclosed in
$r_{2500}$ and $r_{500}$, and also the differential mass fraction as
a function of total mass for our entire sample of 123 systems.
Finally, we address the use of clusters, or more precisely, the total baryon budget and the 
gas mass fraction as a function of redshift, as cosmological tools in Sect.\ref{cosm}.
Given the large scatter observed, we adopt the robust Spearman correlation
coefficient $\rho$ and determine the significance of its deviation from zero $P$ (where 
a small value of $P$ indicates a stronger correlation) to evaluate the significance of the correlations
\citep[see][]{press92}.

\subsection{Cold baryon fraction and star formation efficiency}
\label{cb}

\begin{figure*}[ht]
\centering
\includegraphics[angle=90,width=9.cm]{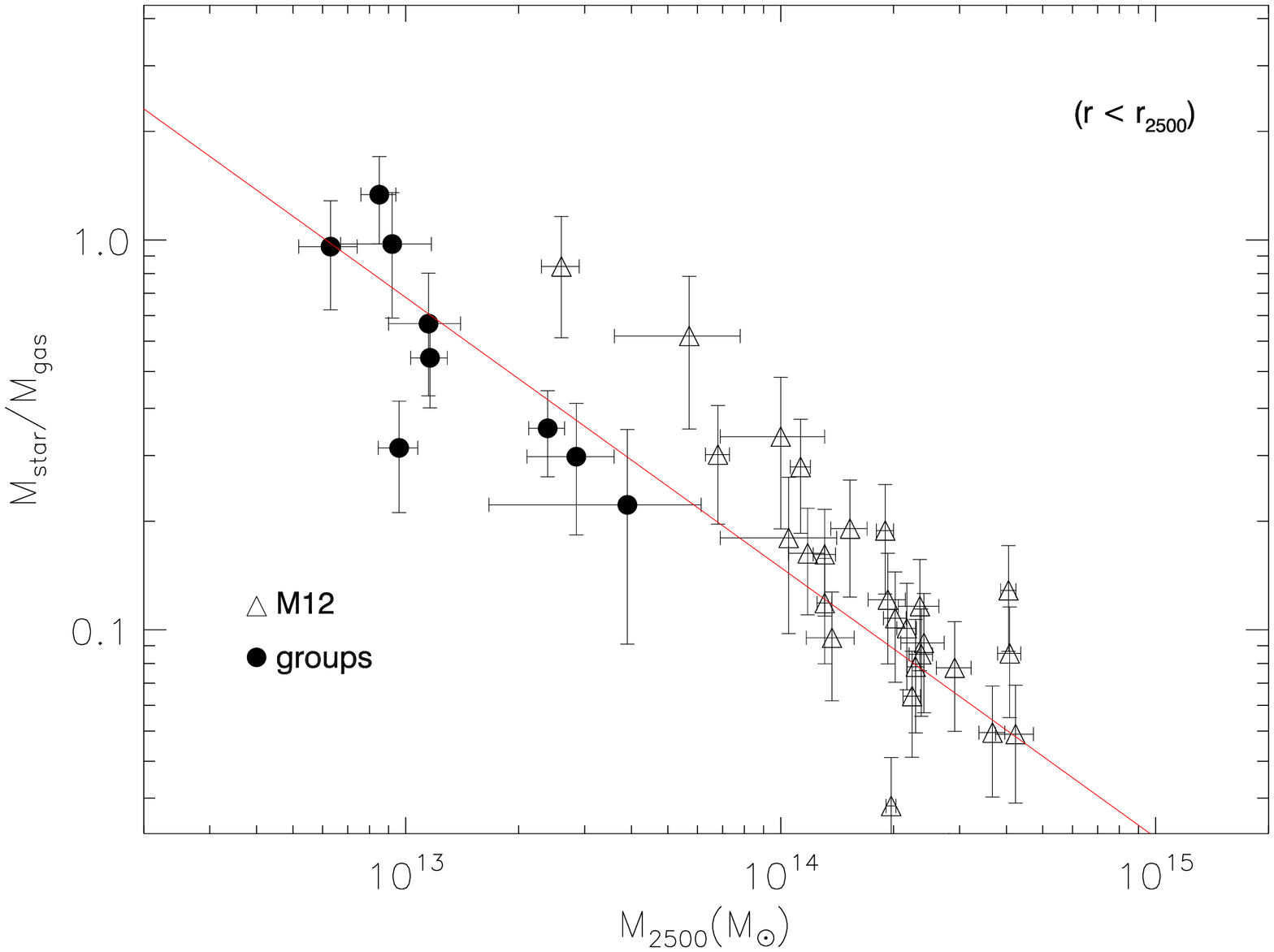}
\includegraphics[angle=90,width=9.cm]{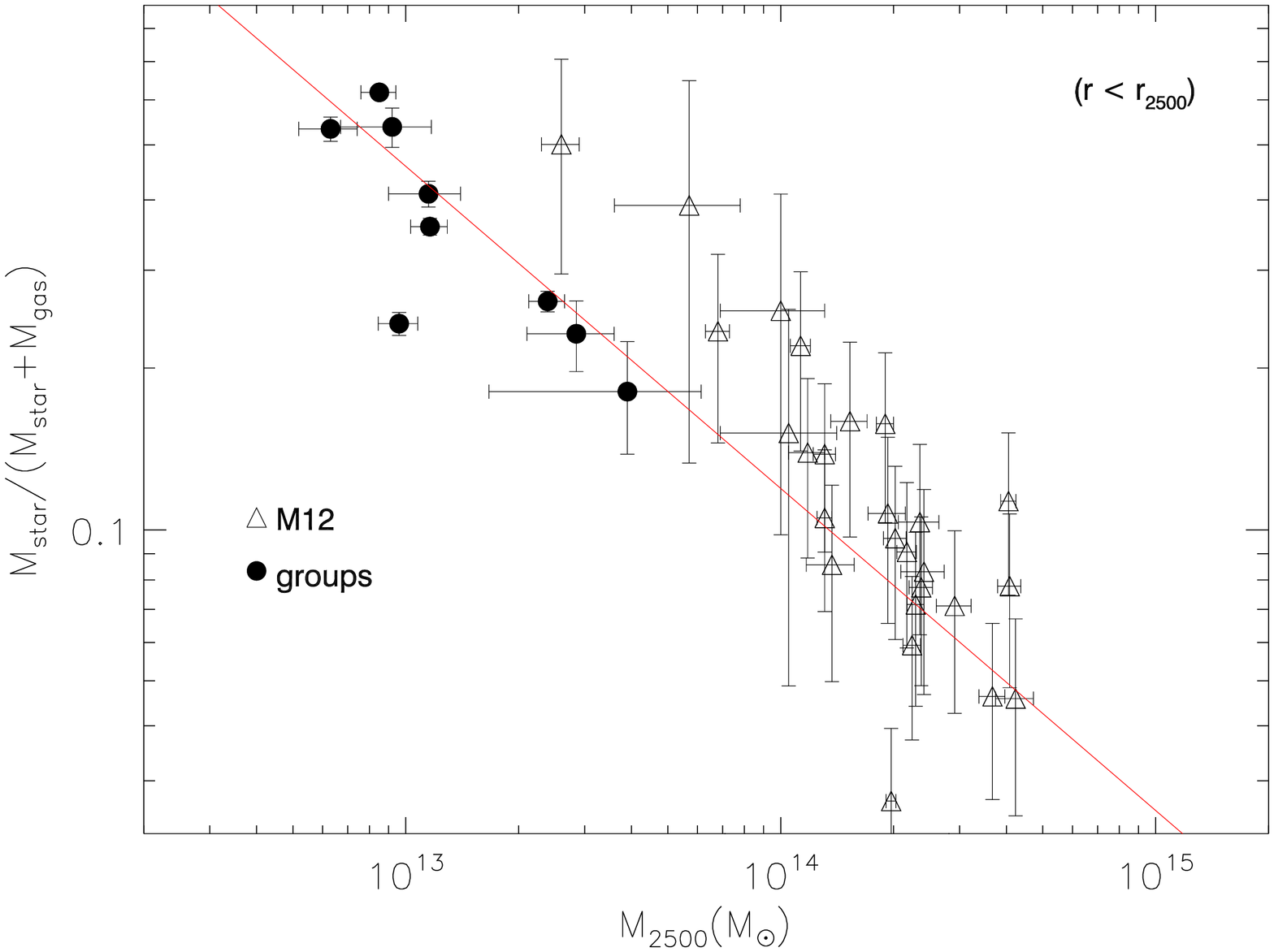}\\
\includegraphics[angle=90,width=9.cm]{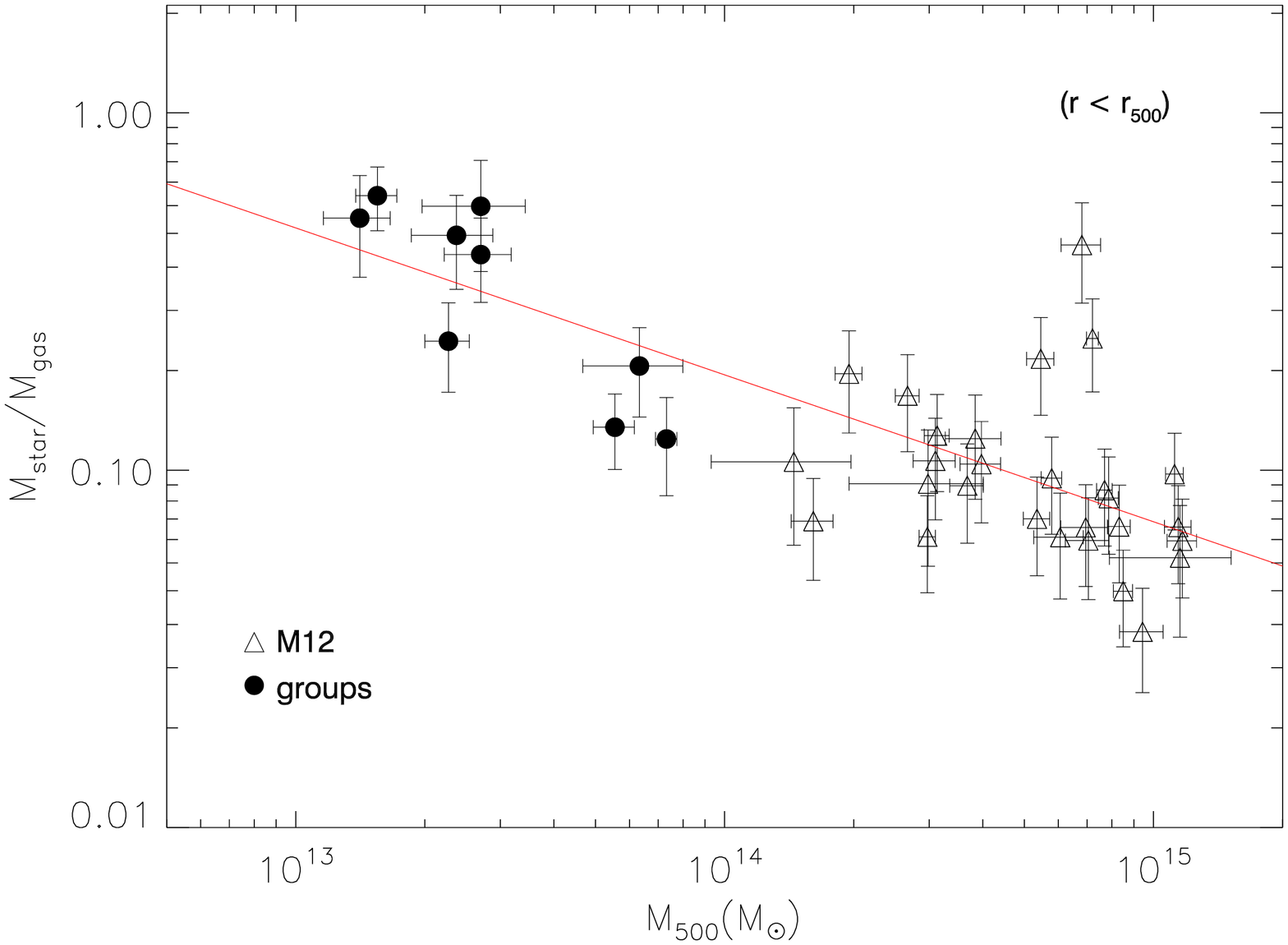}
\includegraphics[angle=90,width=9.cm]{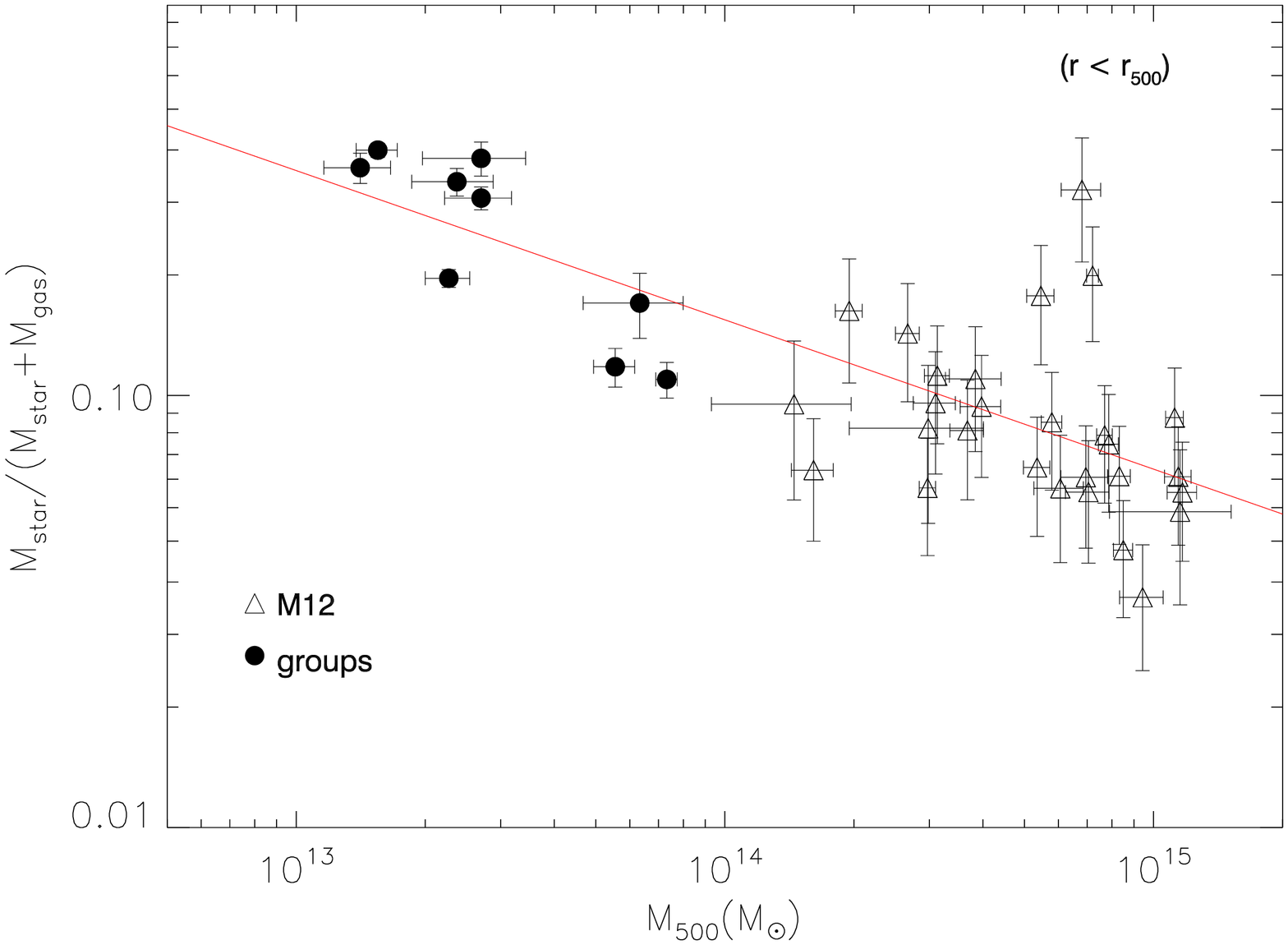}
\caption{Cold baryon fraction, $M_{\rm star}/(M_{\rm gas}+M_{\rm star}$) and star formation efficiency, 
$M_{\rm star}/M_{\rm gas}$, as a function of total mass, computed inside $r_{2500}$ 
(upper panels) and $r_{500}$ (lower panels).}
\label{fig:efficiency}
\end{figure*}

In this section, we investigate the star
  formation efficiency and the cold baryon fraction dependences on
  total mass of the system. The star formation efficiency can be
  defined as the ratio of stellar to gas mass, and the cold baryon
  fraction is the ratio between the stellar mass and the total baryon
  (stars plus ICM) mass.  In Fig.~\ref{fig:efficiency}, we show the
  star formation efficiency and the cold-baryon fraction as a function
  of total mass computed for $r_{2500}$ and $r_{500}$.

From Fig.~\ref{fig:efficiency}, we see a clear trend, which suggests
  that the cold baryon fraction and star formation efficiency decrease
  with increasing total mass \citep[as reported previously
  by][]{david90,roussel00}.  We obtained a Spearman correlation
  coefficient of $\rho=-0.73$ with $P=10^{-7}$ for the correlations
  within $r_{500}$ and $\rho=-0.84$ with $P=10^{-11}$ for the
  correlations within $r_{2500}$. In both cases, we had a strong
  correlation between the cold baryon fraction/star formation
  efficiency and the total mass.

The star formation efficiency decreases by an order of
  magnitude for both $r_{2500}$ and $r_{500}$ from groups to massive
  clusters. From the analysis of twelve groups and clusters,
  \citet{david90} also found a strong correlation for $M_{\rm
    star}/M_{\rm gas} \times  M_{\rm tot}$, showing that the $M_{\rm
    star}/M_{\rm gas}$  ratio varies by more than a factor of
  five from low to high-mass systems.

\subsection{Difference between the observed and WMAP-7 baryon
  fractions}
\label{contr_icl}

\begin{figure}[ht]
\centering
\includegraphics[angle=90,width=9.cm]{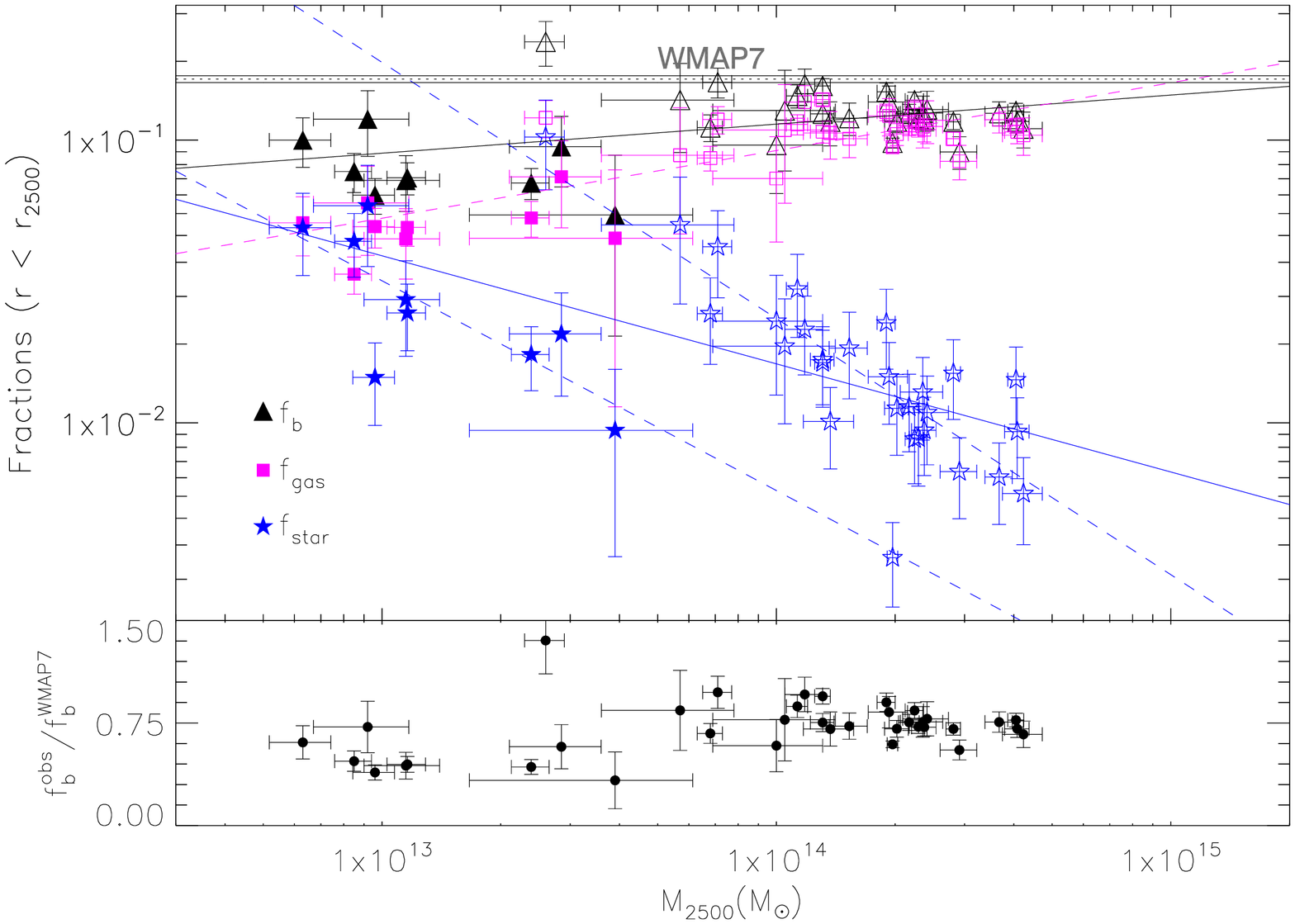}\\
\includegraphics[angle=90,width=9.cm]{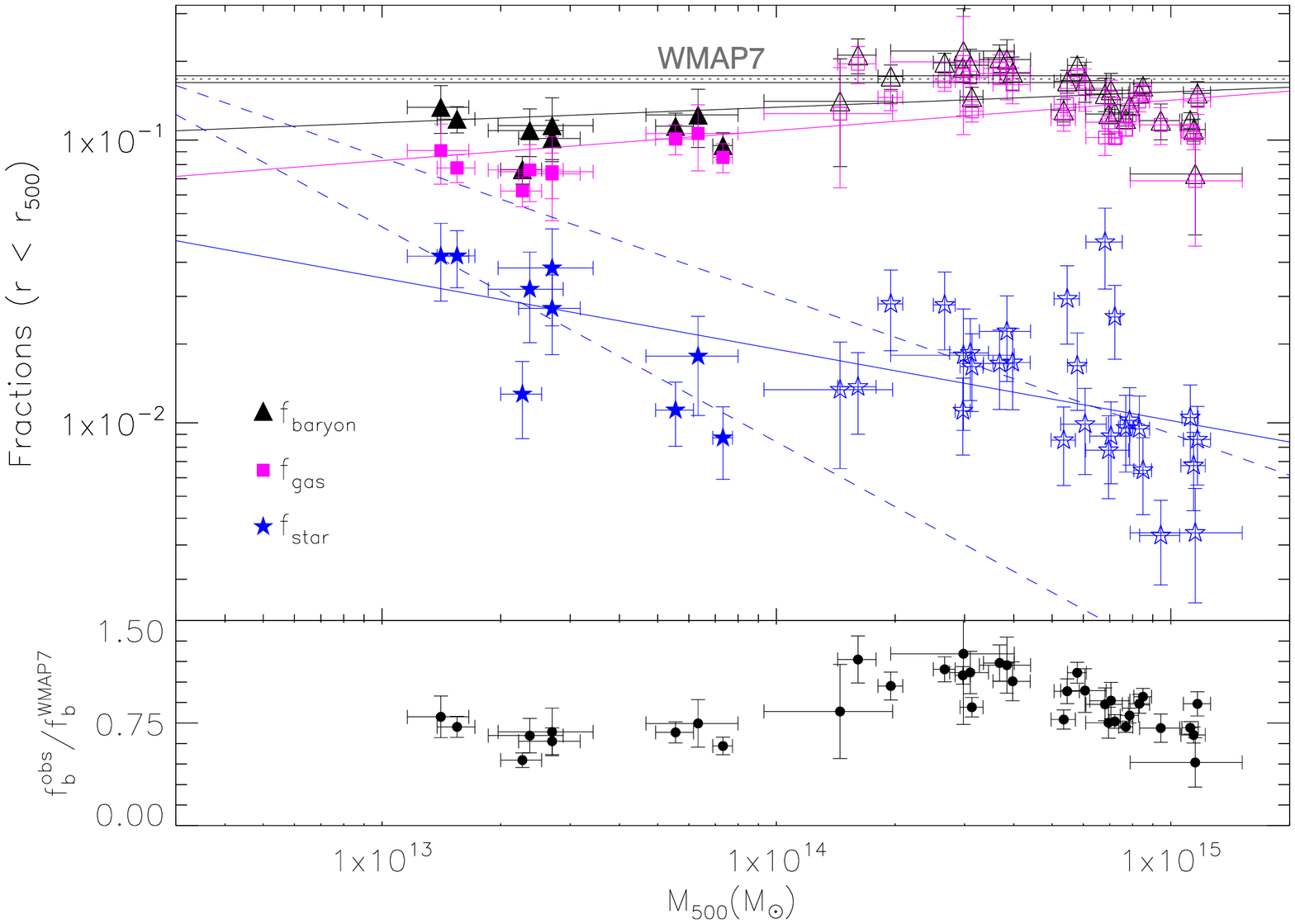}
\caption{Stellar (blue stars), gas (magenta squares), and total baryon (black triangles) mass fractions as a function
of total mass for $r_{2500}$ (top panel) and $r_{500}$ (lower panel). The solid lines correspond to the best linear fit for each
relation. The two blue dashed lines correspond to the fits for the groups and clusters separately.
On the bottom of each panel, we show the ratio between the total baryon fraction determined by the 
sum of $f_{\rm star}$ and $f_{\rm gas}$ and the WMAP-7 value as a function of total mass.}
\label{fig:fracs}
\end{figure}

In Fig.~\ref{fig:fracs}, we show the stellar, gas, and total
  baryon budgets as a function of total mass of the system.  We also
  show the ratio between the total baryon
  fraction determined by the sum of $f_{\rm star}$ and $f_{\rm gas}$
  and the WMAP-7 value as a function of total mass in the bottom panel.  The solid lines
  represent the best linear fits for the total, gas and stellar mass fractions as a function of total
  mass. We also represent the fits for the stellar
  mass fraction  for groups and clusters separately. As we see, the best fits for the
  stellar mass fractions differ by more than $3\sigma$ if we
  separate the 28 clusters from the 9 groups. This suggests
  that groups and clusters form two distinct populations in terms of
  stellar content.

  The total baryon mass fraction, as shown in Fig.~\ref{fig:fracs}, for the
  mass range analysed here indicates a decrease towards groups and
  poor-clusters (as mentioned previously).  The discrepancy between
  the total baryon mass fraction and the WMAP-7 value becomes larger
  with decreasing mass \citep[as already pointed out by some previous
  works, such as in][]{gonzales07,giodini09,andreon10}.  To use the total
  baryon fraction of galaxy clusters as a cosmological tool, one thus should
  consider carefully the sample. These structures range in
  mass from $M_{500} \sim 10^{13} M_{\odot}$ to $10^{15} M_{\odot}$,
  and $f_{\rm baryon}$ is not constant for the entire range of mass.
We discuss this further  in Sect.~\ref{disc}.

\subsection{Gas mass fraction within $r_{500}$, and $r_{2500}$, and 
the differential gas mass fraction ($f_{\rm gas, diff}$)}
\label{fracgas}

The gas mass fraction of clusters of galaxies can be used as a
  cosmological tool, as first done by \citet{allen02} and refined
  later by \citet{allen04} and \citet{allen08}. These studies show an
  agreement between the gas mass fraction and the $\Lambda$CDM
  cosmology. Recently, more sophisticated X-ray analyses were
  done by \citet{ettori09} and \citet{mantz10}.  All these studies
  rely upon the assumption that the gas mass fraction of the cluster
  sample considered is constant with total mass and redshift.

\begin{figure}[ht!]
\centering
\includegraphics[angle=90,width=9.cm]{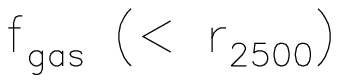}
\includegraphics[angle=90,width=9.cm]{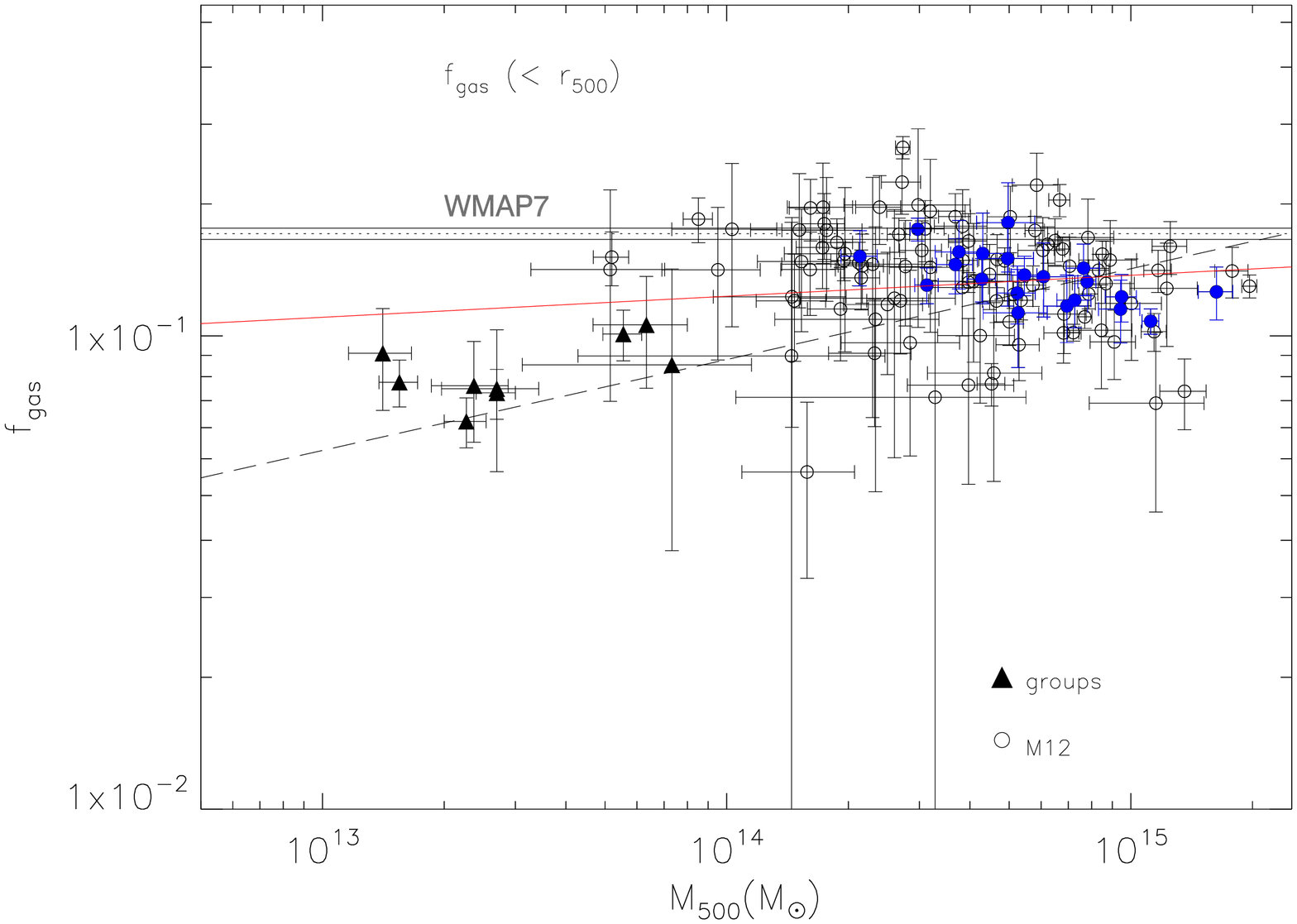}
\includegraphics[angle=90,width=9.cm]{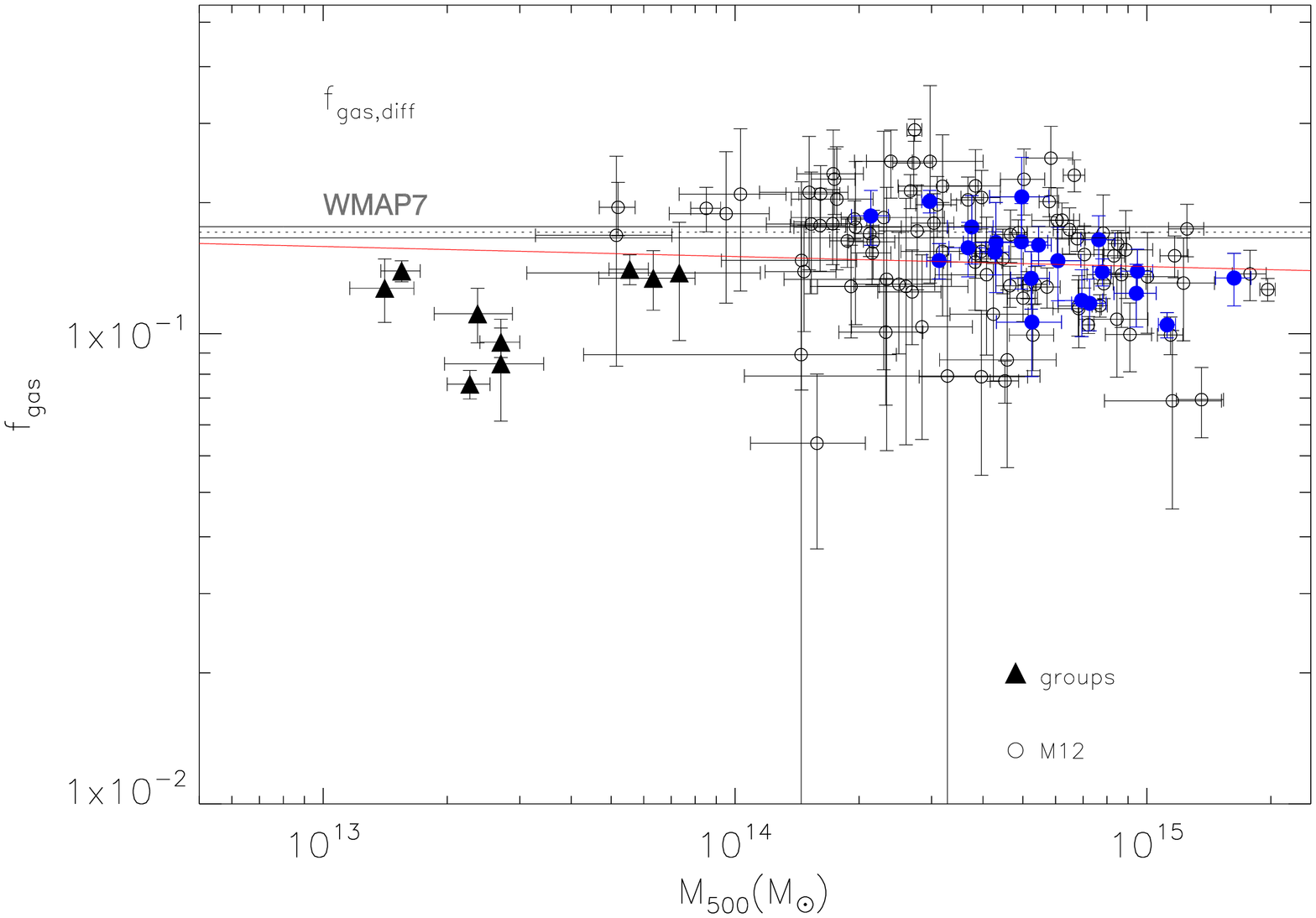}
\caption{Gas mass fraction as a function of total mass computed within
  $r_{2500}$ (upper panel), and within $r_{500}$ (middle panel); 
  the differential gas mass fraction, $[M_{\rm gas,500} - M_{\rm gas,2500}] / [M_{\rm tot,500} - M_{\rm tot,2500}]$,
    as a
  function of $M_{\rm tot,500}$ (lower panel).  The triangles represent the nine groups,
  while the circles represent the 114 clusters from M12 (we
  show in blue the strong cool-core clusters).
  The red
  lines show the best fits for the entire sample. The dashed line in the
  middle panel corresponds to the best fit found by \citet{lagana11}.}
\label{fig:fracgas}
\end{figure}

In this section, we present our results concerning  
the gas mass fraction dependence on the total mass, making use of our entire sample
of 123 objects.  We present our results in Fig.~\ref{fig:fracgas}. We
show the strong cool-core clusters in blue \citep[RCC is defined
in][these are the most relaxed clusters in the sample]{M12}.
Cool-core clusters are generally defined by a drop of
temperature in the centre which is associated with denser cores that
are cooling hydrostatically via bremsstrahlung.  They are associated with
dynamically relaxed clusters, where the hydrostatic equilibrium
equation is valid. However, it seems to be a general agreement 
between the ``true'' masses (measured through weak-lensing) and the
X-ray derived values inside  $r_{\rm 500}$, independent of the dynamical state of the system
 \citep{zhang10}.
 
In Fig.~\ref{fig:fracgas}, we show the gas mass fraction inside
$r_{2500}$, and $r_{500}$ and the differential gas mass
fraction as a function of total mass.  The differential gas mass
fraction, $f_{\rm gas, diff}$, is defined as the mass fraction within the spherical shell of radii $r_{2500}$ and 
$r_{500}$; that is, $[M_{\rm gas,500} - M_{\rm gas,2500}] / [M_{\rm tot,500} - M_{\rm tot,2500}]$.
We see
that there is a  $f_{\rm gas}$ dependence on total mass for total
masses enclosed in both radii.  If we assume a linear
  dependence (BCES bisector), $f_{\rm gas,2500}-M_{\rm tot,2500}$ 
($f_{\rm gas, 2500} \propto M_{2500}^{0.17 \pm 0.02}$ with a $\rho=0.47$ and $P=2.8
\times 10^{-8}$) is compatible with the relation computed for $r_{500}$ within the measurement errors
 ($f_{\rm gas, 500} \propto M_{500}^{0.14 \pm 0.02}$ with a $\rho=0.16$ and $P=0.006$).  
 Both relations are steeper than the relation found for the differential gas mass fraction, 
$f_{\rm gas, diff} \propto M_{\rm tot,500}^{0.02 \pm 0.02}$, that is compatible with a flat distribution.
We must
stress that  the trend found here for $f_{\rm gas,500}-M_{\rm tot,500}$
is not as steep as the one presented in \citet{lagana11} within 1$\sigma$, which may be because
of two major factors: first,  we
computed the total mass from a scaling relation using the gas mass as
a proxy in the previous work, imposing a $f_{\rm gas}$ dependence on $M_{\rm tot}$ and
diminishing the scatter in the relation; second, we have assumed an isothermal gas to compute
the total mass  in the present work, instead of considering the temperature
profile. However, the slopes agree within $2\sigma$.

The differential gas mass fraction shown in Fig.~\ref{fig:fracgas} 
is higher than the cumulative measures, reaching the
universal baryon fraction for systems with mass $M_{500} \sim 10^{14} M_{\odot}$.
Clearly, the differential gas mass fraction is constant at the
1$\sigma$ level and provides a better
constraint for cosmology, although the statistical scatter is very large.
For groups (i.e., the systems in our sample with $M_{500} < 10^{14} M_{\odot}$),
we still observe lower values for the differential gas mass fraction 
when compared to the WMAP-7 result. Since our sample includes few systems in this 
mass regime, we cannot derive
firm conclusions.

Comparing the three plots in Fig.~\ref{fig:fracgas}, we conclude that the
observed $f_{\rm gas}-M_{\rm tot}$ trend found for
groups and clusters is due to the lack of gas enclosed inside $r_{2500}$ for groups and poor clusters, 
as proposed by \citet{sun12}, owing to non-thermal processes,
such as supernova feedback, that are more efficient in low mass
systems because of the shallower potential well. From their hydrodynamical simulations, 
\citet{young11} reported that 
the injection of entropy has removed gas from the cores of the low-mass systems and pushed
the gas out to larger radii between $r_{500}$ and $r_{200}$.

When we analyse the differential stellar mass fractions (i.e.,
$f_{\rm star}$ in the annulus between $r_{2500}$ and $r_{500}$ for the
37 systems for which we have SDSS data) we also do not observe any trend as
a function of $M_{500}$.

\subsection{Cosmological constraints from the total baryon budget and from the $f_{\rm gas}$ vs. redshift relation}
\label{cosm}

As mentioned before, the X-ray data analysis of galaxy
  clusters can provide reliable constraints on cosmology from the total
  baryon budget and the gas mass fraction dependence on the redshift.
  One of the classical methods to infer $\Omega_{\rm m}$ 
  assumes that the  baryon-to-total mass ratio should closely match
  the cosmological values, and thus $\Omega_{\rm b}/\Omega_{\rm m}
  \sim M_{\rm b}/M_{\rm tot}$ \citep{white93,evrard97}. 
Combining $\Omega_{\rm b} = 0.0456$ \citep{jarosik11}
 with our values for the baryon fraction obtained in Sect.~\ref{cb}, 
 our results imply $\Omega_{m} < 0.55$
  \citep[assuming the lowest $f_{\rm b}$ value and agreeing with][]{ettori99}, 
  and if we take the highest
  significant estimates for $f_{\rm b}$, $\Omega_{m} > 0.22$.

In addition to the calculation of $\Omega_{m}$ based on the
total baryon budget, the gas mass fraction has been used to obtain
more rigorous constraints on cosmology that probes the acceleration of
the universe.  The apparent behaviour of $f_{\rm gas}$ 
with redshift can constrain the cosmic acceleration, as studied in 
\citet{ettori99} and \citet{allen04,allen08}. 
This constraint originates because 
$f_{\rm gas}$ measurements depends on the distance to the cluster ($f_{\rm
gas} \propto d^{1.5}$).

\begin{figure}[ht!]
\centering
\includegraphics[angle=90,width=9.cm]{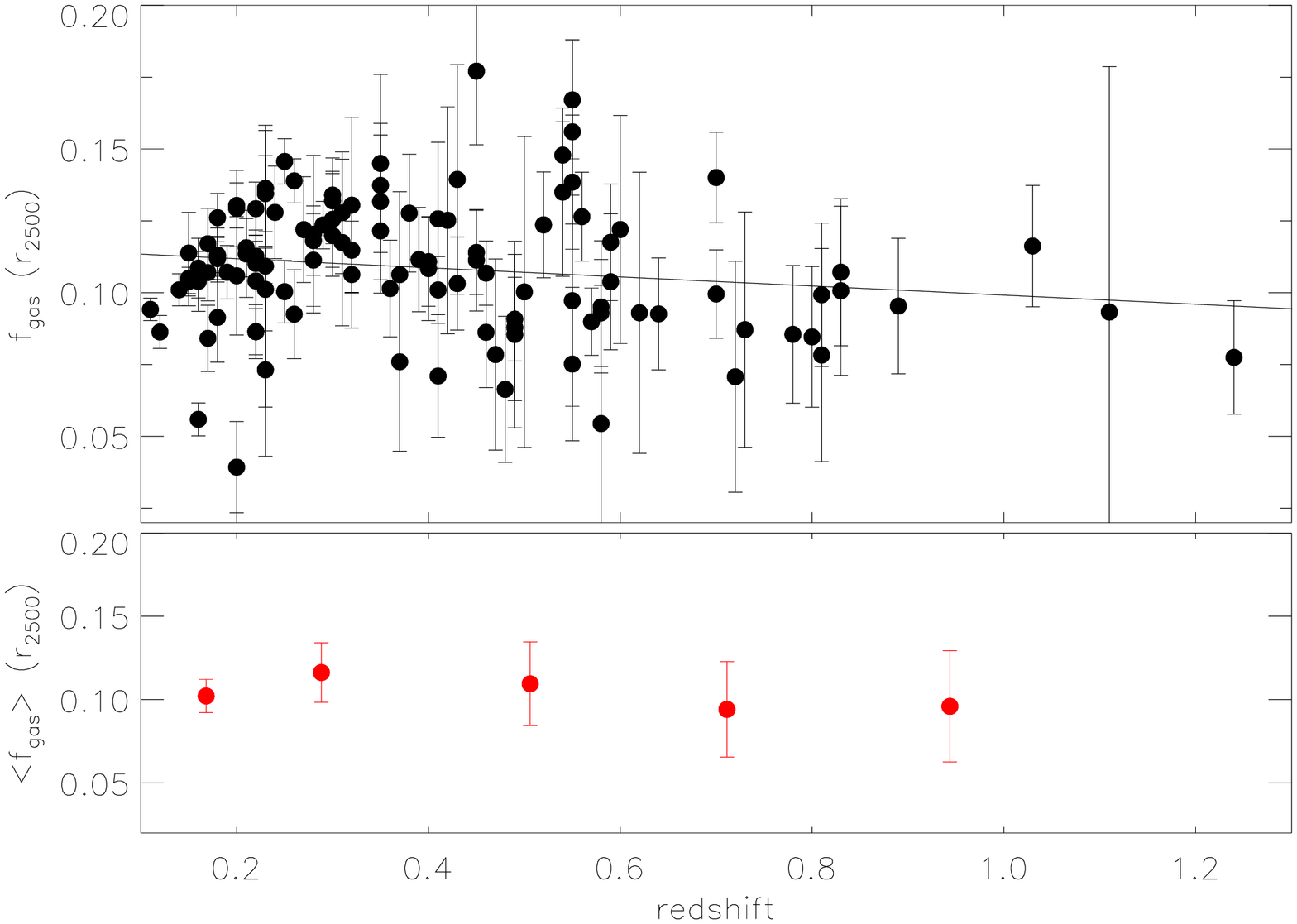}
\includegraphics[angle=90,width=9.cm]{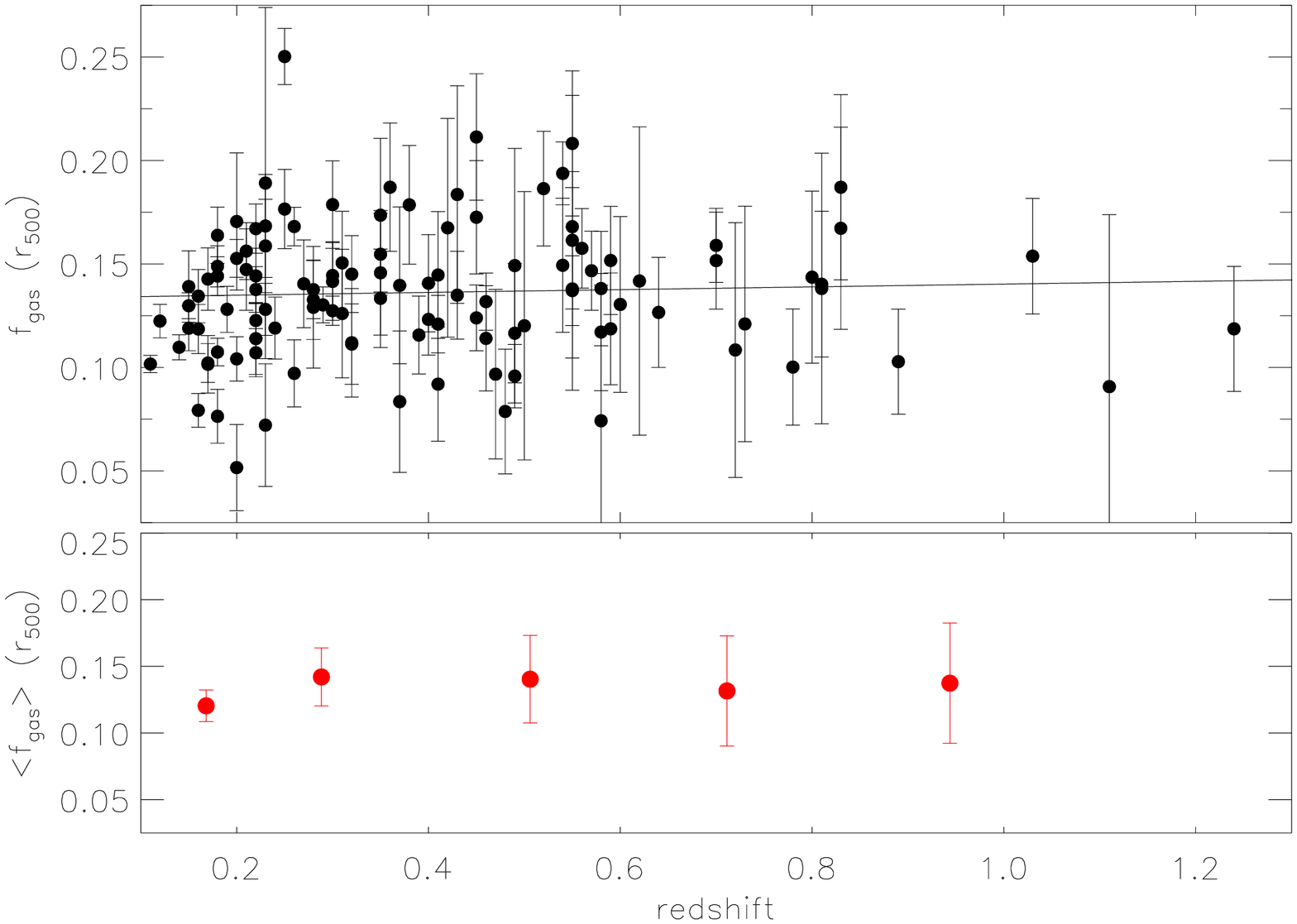}
\caption{Gas fraction as a function of redshift for $r_{2500}$ (upper panel) and $r_{500}$ (lower panel). 
The black lines represent the best linear fit for the data.}
\label{fig:fgasz}
\end{figure}

The gas mass fraction for the present sample was computed,
  assuming the default $\Lambda$CDM cosmology, and we show the 
  gas mass fraction determined inside
  $r_{2500}$ ($f_{\rm gas,2500}$) and $r_{500}$ ($f_{\rm gas,500}$)
  as a function of redshift in
  Fig.~\ref{fig:fgasz}. On the bottom of each panel, we show 
  the mean gas mass fraction value computed for bins of 0.2
  in redshift as a function of the mean redshift with
  red points, and the errors are
  the standard errors on the mean.

From this figure, we can see that the gas mass fraction values
  determined for $r_{2500}$ and $r_{500}$ show little variation with
  redshift. Assuming $f_{\rm gas(z)} = A + B \times z$, we obtained
  $A=0.130 \pm 0.009$ and $B=0.010 \pm 0.015$ for $r_{500}$, and
  determined $A=0.110 \pm 0.008$ and $B=-0.019 \pm 0.013$ for $r_{2500}$.  From
  these results, we verified that the enclosed gas mass fraction does
  not depend on redshift at a $2\sigma$ level in both cases. 

From our analysis, we cannot attribute the slight decrease of $f_{\rm
    gas,2500}$ with redshift to evolution of $f_{\rm gas}$.  Selection
  biases may have led to the enhanced mean gas mass fraction at
  around z=0.3 and may have driven the observed trend, as
  recently reported by \citet{Landry12}.

\section{Discussion}
\label{disc}

In this section, we further discuss the results obtained in this work
and compare them to previous observational and theoretical ones.  

\subsection{Stellar and gas mass dependence on total mass}

 The trends shown in Fig.~\ref{fig:efficiency} can be
  explained by a decrease in the stellar mass with an increase in cluster
  mass. Alternatively, the gas mass can increase for more massive
  systems.  From previous results in the literature
  \citep{david90,lms03,lagana08,giodini09,lagana11,zlp11}, we observe
  both behaviours, but the decrease in stellar mass fraction as a
  function of total mass is more significant than the gas mass
  increase in the same range. This provides evidence that there is no
  well-tuned balance between the gas and stellar mass fractions.  This
  evidence suggests that the variation in star
  formation efficiency is because of the variation in stellar mass, which arises from
  the decrease in efficiency of tidal interactions among galaxies, the
  removal of the gas reservoir of galaxies due to the motion of galaxies through the ICM, 
  or feedback processes that may quench star-formation.
Recent numerical simulation results from \citet{Dubois13} supported this scenario.
According to their results, AGN feedback alone is able to significantly alter the stellar mass content
by quenching star formation.

During mergers to form rich clusters, the gas within the
  system will be shocked and heated to the virial temperature.  As
  mergers progress, more massive systems are formed, and the gas is
  progressively heated to higher temperatures. Thus, cooling and
  galaxy formation are inhibited. Since rich clusters are formed from
  many mergers, this explains why this large gas fraction of the gas
  is not consumed to form stars.  In contrast, less massive
  systems, such as groups, have experienced fewer mergers. Therefore their
  gas is cooler and more stars are formed within the galaxies. As a
  consequence, the star formation efficiency is higher in groups, which
  indicates that physical mechanisms that depend on the virial mass, such
  as ram-pressure stripping,  are driving galaxy evolution within
  clusters and groups.  This result is important to provide
constraints on the role of thermodynamical processes for groups and
clusters and seems to agree with theoretical expectations
from hydrodynamical simulations by \citet{SH03}.
In their models, the integrated star formation efficiency as a function of halo mass, which varies from $10^{8}$ 
to $10^{15} M_{\odot}$, falls by a factor of five to ten over the cluster mass scale, due to
the less efficient formation of cooling flows for more massive haloes (in their case, with temperature above 
$10^{7}$ K, what comprises the entire range of mass analysed in this work).

Recently, \citet{planelles13}  carried out two sets of simulations including radiative cooling, 
star formation and feedback from supernovae and in one of which they also accounted for the effect of
feedback from AGN. These authors found that both radiative simulation sets
predict a trend of stellar mass fraction with cluster mass that tends to be weaker
than the observed one. However this tension depends on the particular set of
observational data considered. Including the effect of AGN feedback alleviates
this tension on the stellar mass and predicts values of the hot gas mass fraction
and total baryon fraction to be in closer agreement with observational results.
Also, \citet{zehavi12} studied the evolution of stellar mass in galaxies as a function of host halo mass using semi-analytic
 models, and their results agree with our findings of a varying star formation efficiency. These latter authors found that baryon 
 conversion efficiency into stars has a peaked distribution with halo mass and that the peak location shifts toward lower mass from 
 z$\sim$ 1 to z$\sim$ 0. Another difference between low- and high-mass haloes is that the stellar mass
 in low-mass haloes grows mostly by star formation since z$\sim$1. In contrast,  most of the stellar mass is assembled by mergers
 in high-mass haloes.

%MORE SIMULATION RESULTS

\subsection{The ICL contribution to the total baryon budget}

To account for the difference in baryons at the low-mass end (see
Fig.~\ref{fig:fracs}), where the total baryon budget is still
significantly below the value of WMAP-7, there are two
possibilities. The first possibility is that the discrepancy in
baryon mass fraction is because of  feedback mechanisms, as expected by many studies
\citep[including][]{bregman07,giodini09}. Recent numerical
simulations performed by \citet{dai10} proposed that the baryon loss
mechanism is primarily controlled by the depth of the potential well:
the baryon loss is not
significant for deep potential wells (rich clusters), while for lower-mass clusters 
and groups the baryon loss
becomes increasingly important. Baryons can be expelled from the
central regions beyond $r_{500}$.  The second possibility is to
account for baryons in other forms: for example, the increase in the
stellar light via intra-cluster light (ICL) which increases the stellar
mass.

ICL is one of the most important sources of
unaccounted baryons, and observational results have shown that it
may represent from 10\% to 40\% of the total cluster light  
\citep[e.g.,][]{feldmeier02, zibetti05, kb07, gonzales07}.
However, recently, \citet{burke12} found that  the ICL constitutes only 1-4\% of the total
cluster light within $r_{\rm 500}$ in high-z clusters.

To investigate the ICL contribution as a function of system total mass, we computed 
the difference between the observed total baryon budget and the WMAP-7 value, which is taken to be:
$ M_{\rm diff}= (f_b^{WMAP-7}-f_b^{obs}) \times M_{500},$
where $f_b^{WMAP-7}$ is the WMAP-7 value; $f_b^{obs}$ is the observed
baryon mass fraction that we computed for our
systems; and $M_{500}$ is the total mass computed in the $r_{500}$
radius.
We then assumed that this difference in mass is under the form of
luminous matter. To compute  the corresponding ``missing stellar surface
brightness'', we assumed a mass-to-light ratio for ellipticals of $M/L= 1.7 
~M_{\odot}/L_{\odot}$ \citep{kauff03}  to convert the difference in
mass into a difference in luminosity. We then convert the luminosity in
surface brightness by dividing the area inside of $r_{500}$.

\begin{figure*}[ht]
\centering
\includegraphics[angle=90,width=7.cm]{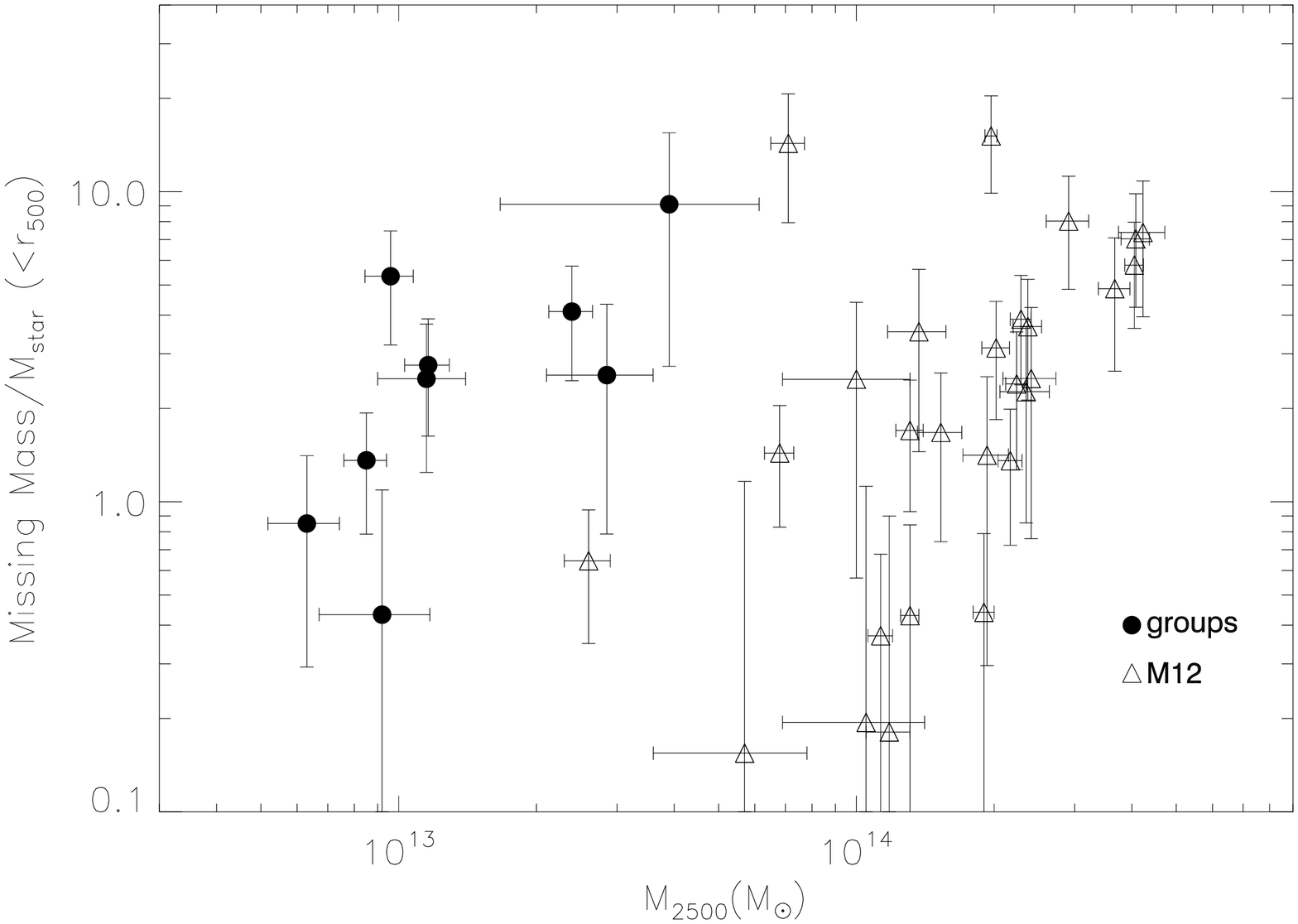}
\includegraphics[angle=90,width=7.cm]{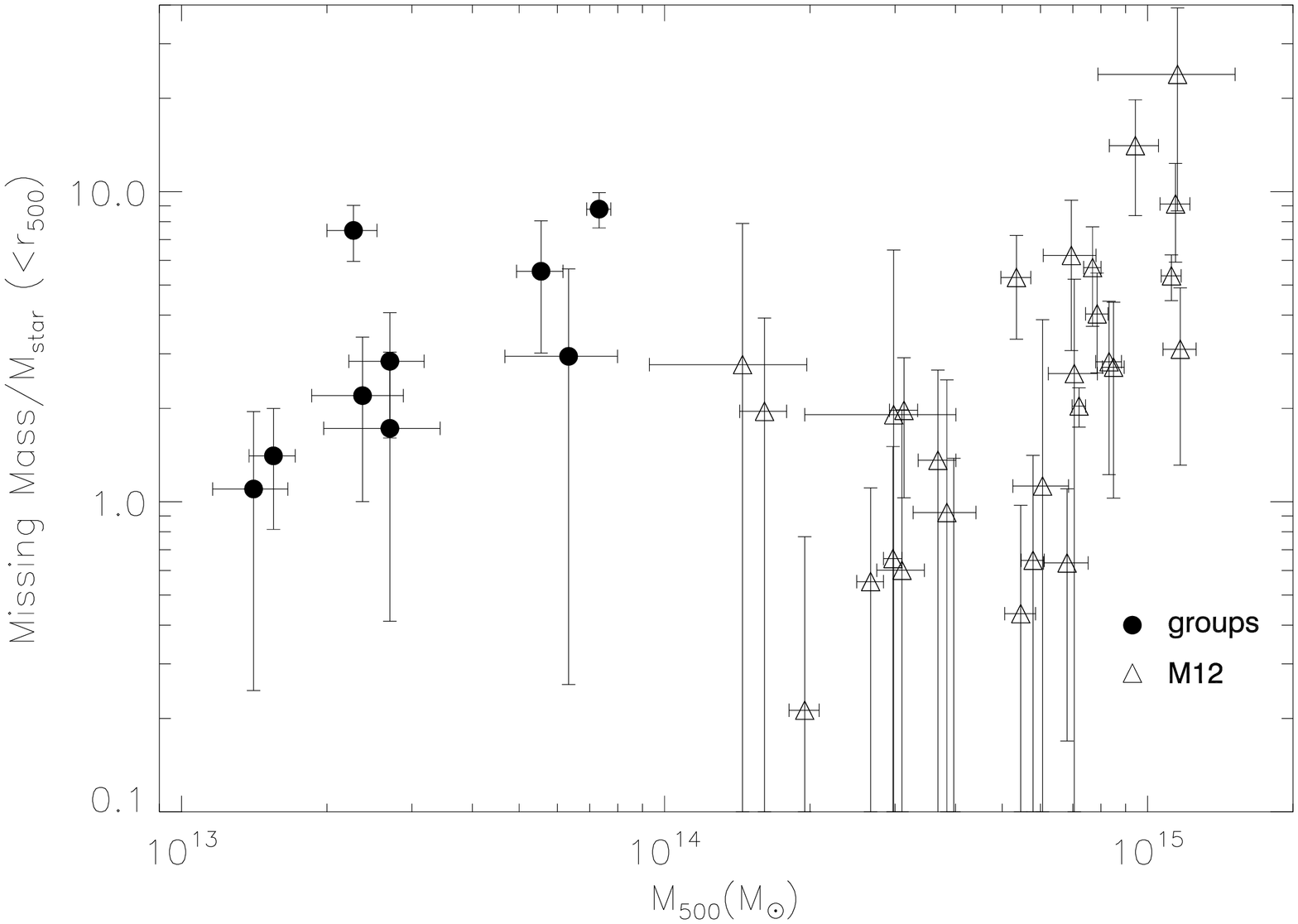}\\
\includegraphics[angle=90,width=7.cm]{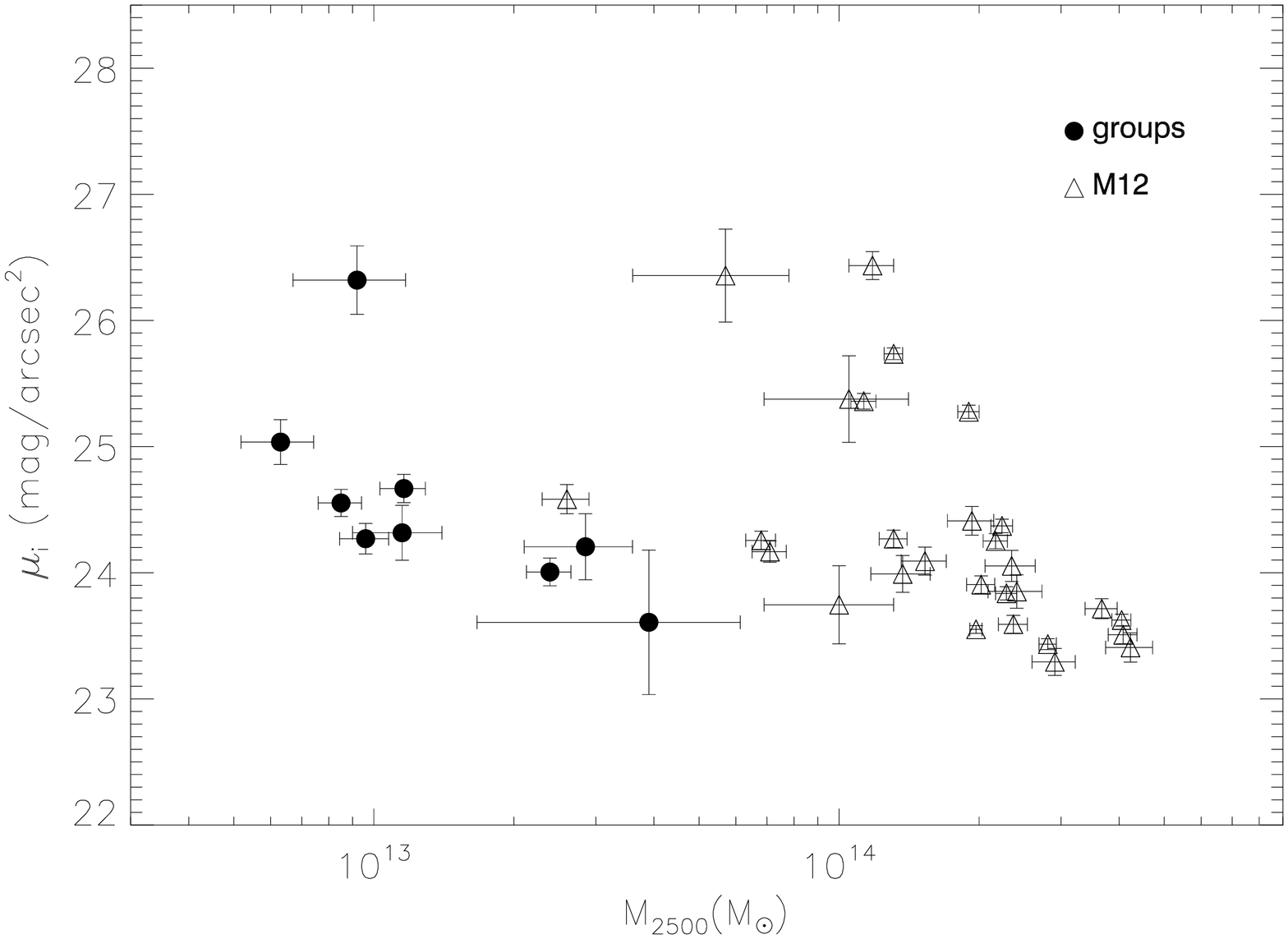}
\includegraphics[angle=90,width=7.cm]{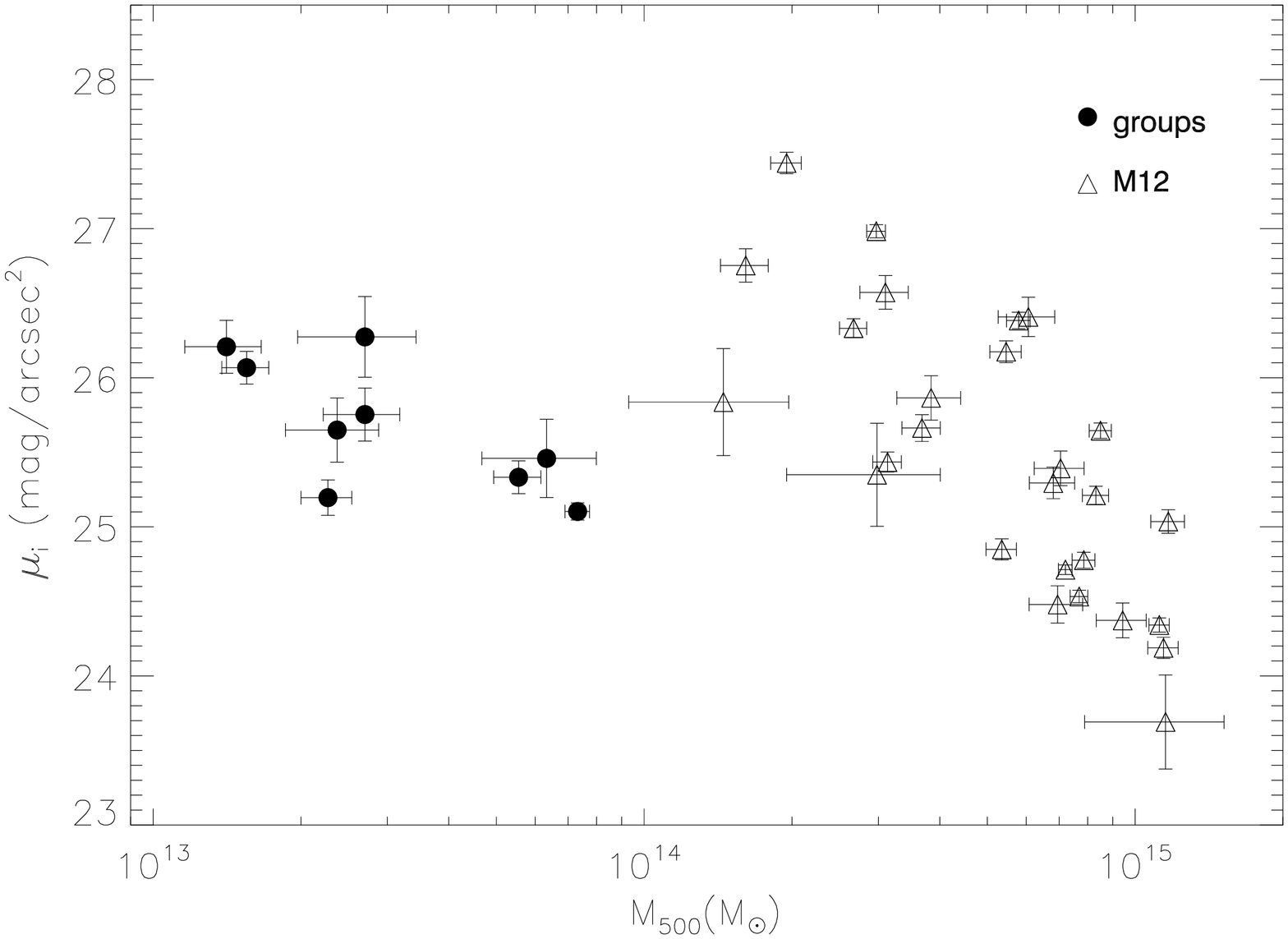}\\
\includegraphics[angle=90,width=7.cm]{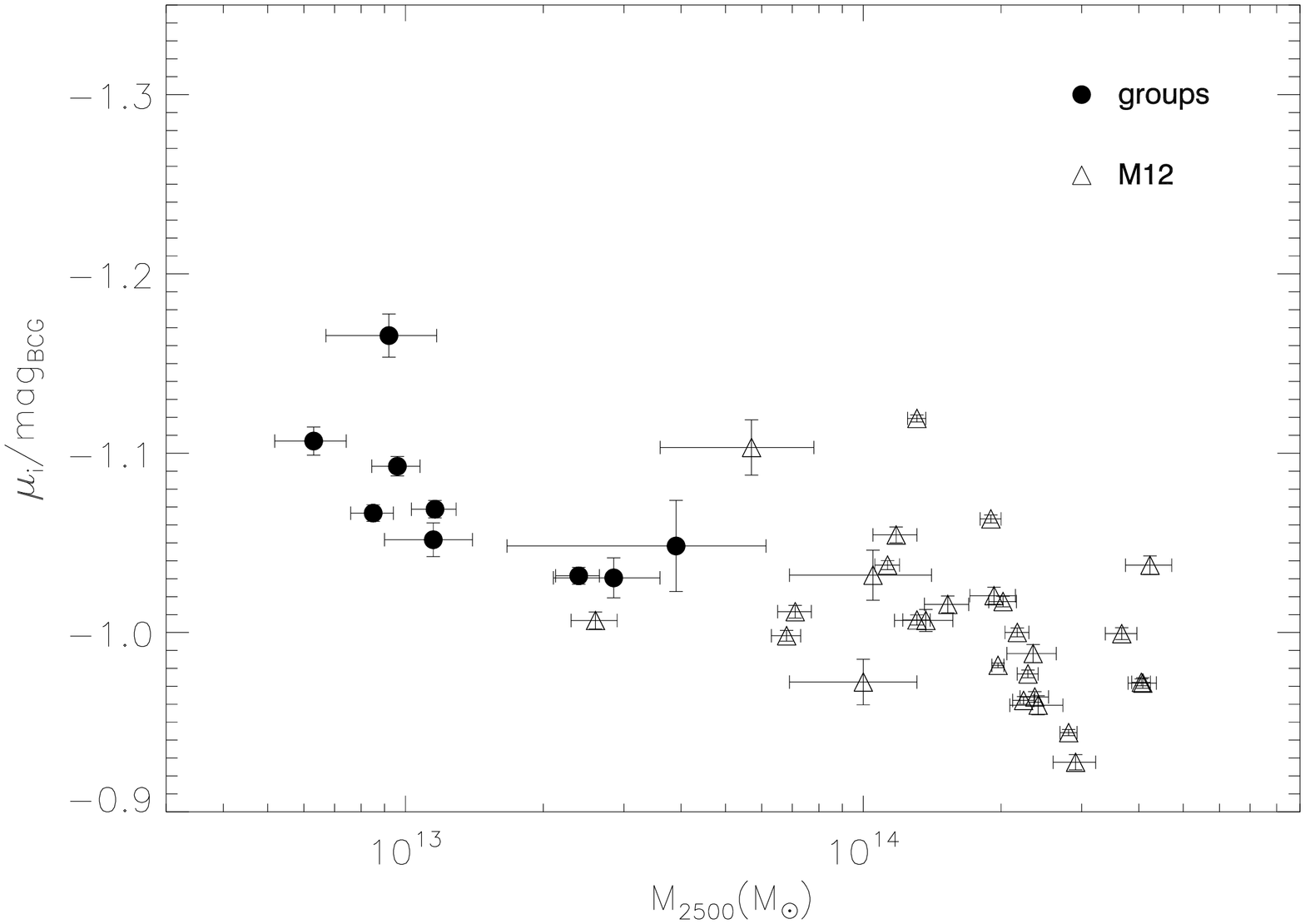}
\includegraphics[angle=90,width=7.cm]{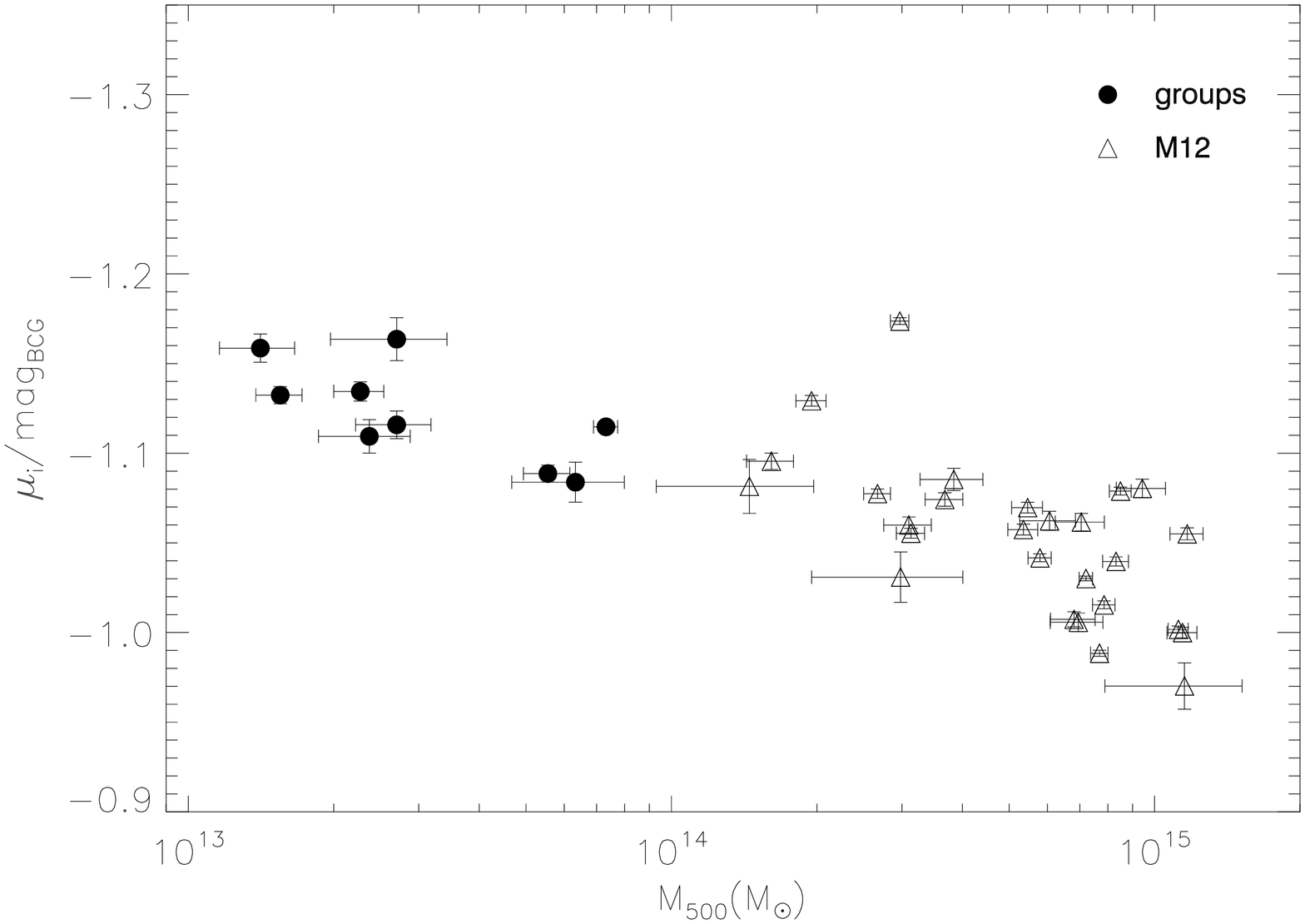}
\caption{\textit{Top}: ``missing stellar mass'' to total stellar mass ratio as a function of $M_{2500}$ (left panel) and $M_{500}$ (right panel);
   \textit{Middle}: ``missing stellar surface brightness'' as a function of total mass for $r_{2500}$ (left panel) and $r_{500}$ (right panel);
   \textit{Botton}: ``missing stellar surface brightness'' to BCG magnitude ratio as a
  function of $M_{2500}$ (left panel) and $M_{500}$ (right panel).}
\label{fig:missmass}
\end{figure*}

In Fig.~\ref{fig:missmass}, 
we show 
the missing stellar mass divided by the total stellar mass 
as a function of total mass (upper panel),   the missing stellar surface brightness as a function of
total mass (middle panel), and the ratio between the stellar surface brightness and the BCG magnitude
as a function of total mass. Since the stellar mass fraction
decreases for massive clusters (Fig.~\ref{fig:fracs}) one could conclude that the 
ICL could be more important in clusters of galaxies because the measured stellar light fraction of
the ICL has not been considered. 
As a direct consequence of the stellar mass dependence on total mass, we observe an increase 
of the missing-to-total stellar mass ratio
 as a function of the total mass of the system,
as shown in  the upper panel of Fig.~\ref{fig:missmass}.
In this panel, groups and  clusters seem to be two distinguished populations, 
which are comparable to the behaviour observed in  Fig.~\ref{fig:fracs}.
Then, if we think in terms of missing surface brightness, groups seems to present
fainter values than clusters. This means that it would be easier to detect the ICL contribution is 
 clusters than in groups.

However, from the values obtained here, the stellar component would have to increase by a factor of three 
to be able to explain the baryon deficit with intra-cluster light for low-mass systems. Moreover, such a high amount of ICL
would be visible in current observations. Although a large sample of clusters and groups is needed for further constraints 
on the photometrical properties and for the ICL formation mechanism, it is very unlikely that the ICL will
be able to answer the baryon deficit problem.

Even by considering the uncertainties associated with our total baryon
fraction and the WMAP-7 measurement, the two values are
discrepant at the  $5\sigma$ level for systems with mass $M_{500} <
10^{14}M_{\odot}$. It is therefore probable that either
 the baryon mass fraction within $r_{500}$
in groups is different from the universal value, or that there are still
unaccounted baryons (other than ICL) in the low-mass regime.

\section{Conclusions}
\label{conc}

We have investigated the baryon content of a sample of 123 galaxy
clusters and groups, whose mass spans a broad range (from groups of
$M_{500} \sim 10^{13} M_{\odot}$ up to clusters of $M_{500} = 4 \times
10^{15} M_{\odot}$). We measured the cluster gas and total
masses from X-ray data analysis, and we computed the stellar masses for
37 out of 123 galaxy clusters and groups using DR8-SDSS data. 
We summarise our results and
discussions in the following points:

\begin{itemize}

\item{For a subsample of 37 systems for which we had both optical and
    X-ray data we derived quantities inside $r_{2500}$ and $r_{500}$ 
     to investigate the stellar, gas and total mass fractions dependence on total mass. 
     We confirmed the previous trend found in the literature: the
    star formation efficiency is lower for more massive clusters.
    It decreases by an order of magnitude from groups to clusters
    inside both  $r_{2500}$ and $r_{500}$. We observe a decrease of the cold baryon 
    fraction and of the star formation efficiency
    from $r_{2500}$ to $r_{500}$.}

\item{Star-formation efficiency is lower in galaxy clusters
      than in groups, which suggests that the gas reservoir of the galaxies during cluster formation 
      in more massive clusters are more affected by mechanisms, such as ram-pressure, that quench the star formation.
   Physical mechanisms depending on the total mass of the system may be driving galaxy evolution in groups and
      clusters. This observational result agrees with
      hydrodynamical simulations.}

\item{For the entire sample of 123 systems, we analysed the
       gas mass fraction inside $r_{2500}$, and $r_{500}$
      and also the differential gas mass fraction ($f_{\rm gas, 2500}$
      - $f_{\rm gas, 500}$) dependence on  total mass.  We found that the gas fraction depends
      on the total mass inside both radii, with the dependence being
      steeper for the inner radius. However, we found that for systems more massive than
      $M_{\rm tot} > 10^{14} M_{\odot}$, the differential gas mass fraction shows no 
      dependence on total mass, which provides evidence
      that groups cannot retain gas in the inner parts in the same way
      clusters do.  Since groups have shallower potential wells,
      non-thermal processes are more important than in clusters, and
      AGN feedback, for instance, could expel the gas towards the
      outer radii of groups. This result is an indication that such
      processes must play a more important role in the centres of low-mass
      systems than in massive clusters.}

\item{The differential gas mass fraction is higher than the
      cumulative measures and clearly more constant as a function of total mass, which provides better
      constraints for cosmology.}

\item{Our results show that non-thermal processes play
      different but important roles on galaxy evolution in groups and
      clusters. While in groups, the gas is more affected because of the lower
      potential well, in galaxy clusters, the cold baryons (stars in
      galaxies) are more affected due to ram-pressure stripping.
      These mechanisms make the baryon distribution in these two
      structures  different.  Many studies have indeed reported
      systematic differences between the physical properties of groups
      and the clusters.}

\item{To investigate the contribution of the ICL to the baryon
      budget, we computed the difference in mass between the WMAP-7
      value and the observed baryon fraction in terms of surface
      brightness. 
      The values found here for
      the expected missing surface brightness for clusters of galaxies are
      similar to the ICL surface brightness detected in the
      literature, which span the range of $\sim 24 < \mu_{\rm r} < 28$.
      If the difference between the observed baryon fraction and the WMAP-7 value is because of luminous
      matter that is spread throughout the volume of the system, 
      as a direct consequence of $M_{\rm star}$ dependence on $M_{\rm tot}$,
      we observe a small increase in Missing Mass/$M_{\rm star}$ with total mass,
      and groups and clusters seem to be separated into two distinct populations.  
      However, we have a high scatter inside both $r_{2500}$ and $r_{500}$.
      In terms of missing surface brightness, we observe a 
      small decrease as a function of total mass also for the 
       ratio between the missing surface brightness and the BCG
      magnitude inside $r_{500}$, but it is difficult to spot a clear trend from these panels.}
    
\item{Combining the baryon-to-total mass fraction with
      primordial nucleosynthesis measurements, our results indicate
      that $0.17 < \Omega_{\rm m} < 0.55$. We also observed that
      the gas mass fraction that is enclosed within $r_{2500}$
      and $r_{500}$ does not depend on redshift at the 2-sigma level, which is very
      important in the use of clusters of galaxies as a cosmological tool to
      constrain the cosmic acceleration.}

\end{itemize}

\begin{acknowledgements}
We are very grateful to Christophe Adami and Emmanuel Bertin for their 
advice during the early stages of this work. We are also grateful to the anonymous
referee who carefully read this paper and gave important suggestions.
We acknowledge financial support from CNES and CAPES/COFECUB program 711/11.
T. F. L acknowledges financial support from FAPESP (grant: 2008/0431-8).
Y.~Y.~Z. acknowledges support from the German BMWi through the Verbundforschung under the
grant NO \,50\,OR\,1103.

\end{acknowledgements}

\begin{appendix}
\label{app}
\section{Testing the isothermal assumption to compute $r_{500}$ and $M_{\rm tot, 500}$}

To test if the assumption of an isothermal cluster would
introduce any systematic effect on the total mass determination, 
biasing low the total mass of cool-core clusters, we consider eight cool-core
clusters from \citet{V06} for which all the necessary parameters were
available.  We computed $r_{500}$, $M_{\rm tot, 500}$ and $M_{\rm gas,
  500}$ by considering their Eq.~(6) to describe the temperature profile
and a mean temperature,  $T_{\rm mg}$, in the 
isothermal case.  For both cases, the emission measure profile is given by
the sum of a modified $\beta$-model profile and a second $\beta$-model
component with a small core radius, as stated in their Eq. (2).

In Fig.~A.1, we show the comparison between $r_{500}$, $M_{\rm tot,
  500}$ and $M_{\rm gas, 500}$ which is computed in both ways,
  using a temperature profile and assuming an isothermal case.  
To compute the
gas mass, we do not use the temperature profile, since the values of
$r_{500}$ change. The enclosed $M_{\rm gas, 500}$ will also change,
and  for completeness we thus, show the comparison for the gas masses computed in both ways
in Fig.~A.1.

\begin{figure}[ht!]

\centering
\includegraphics[width=9.cm]{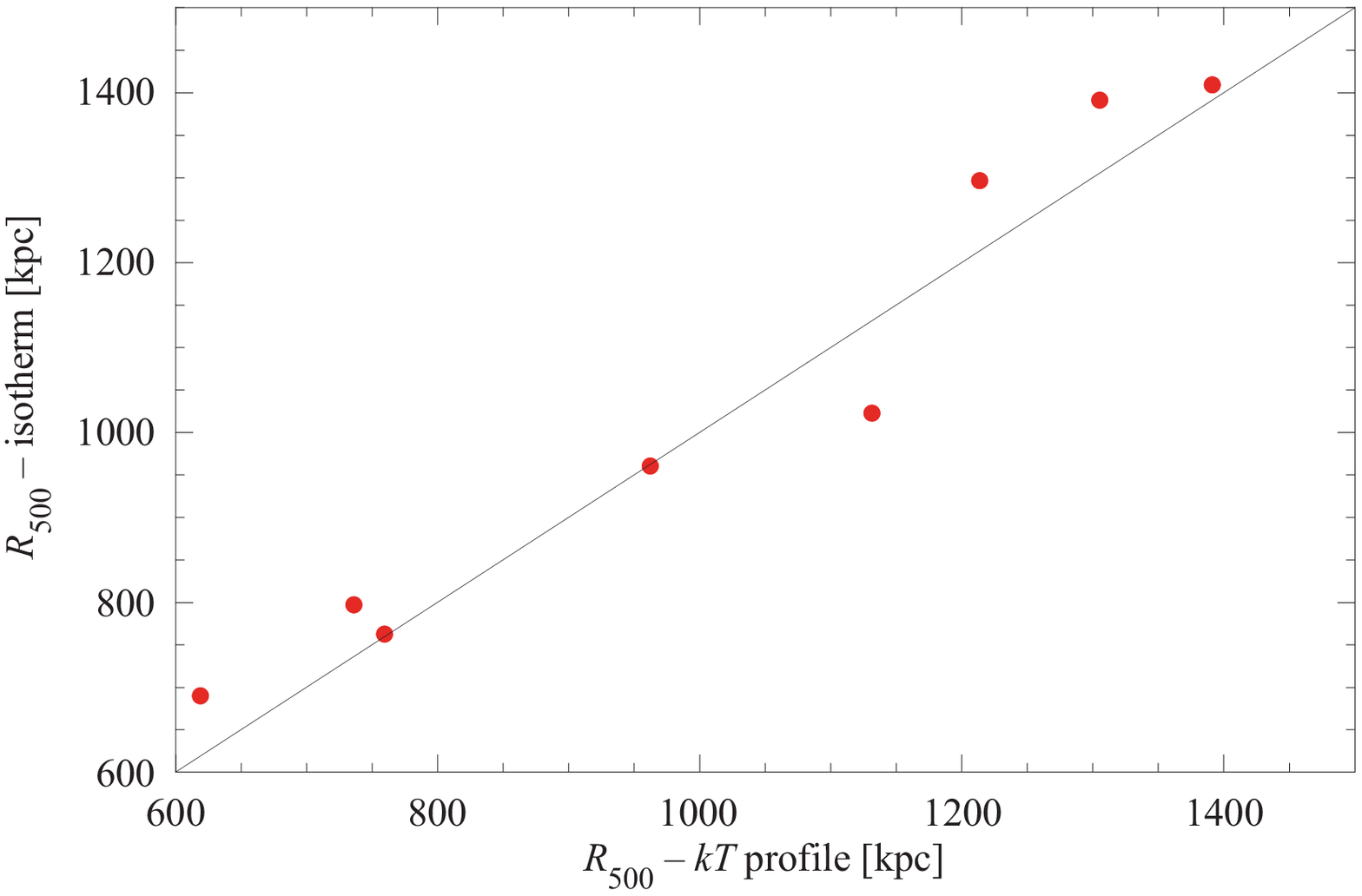}
\includegraphics[width=9.cm]{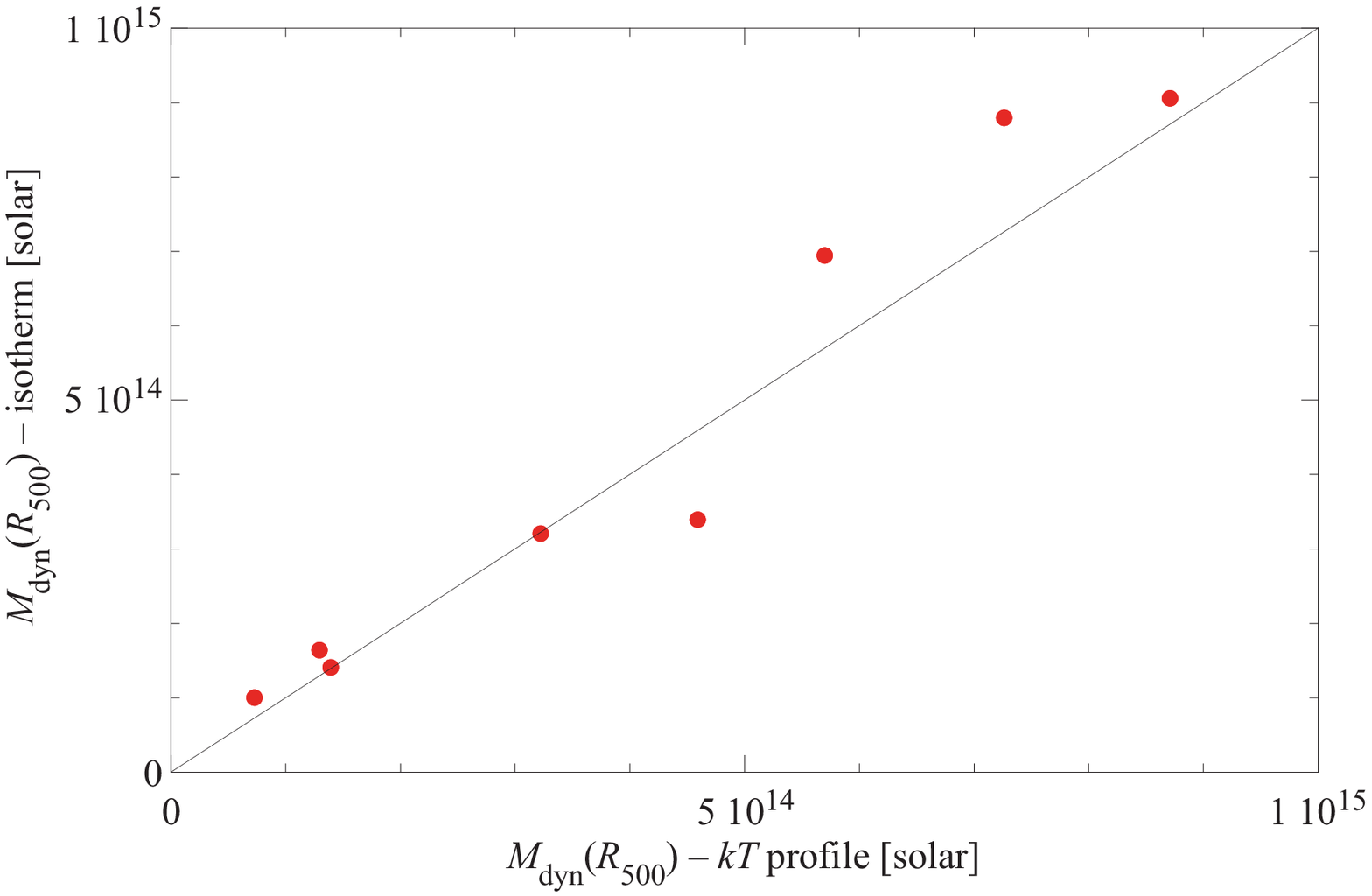}
\includegraphics[width=9.cm]{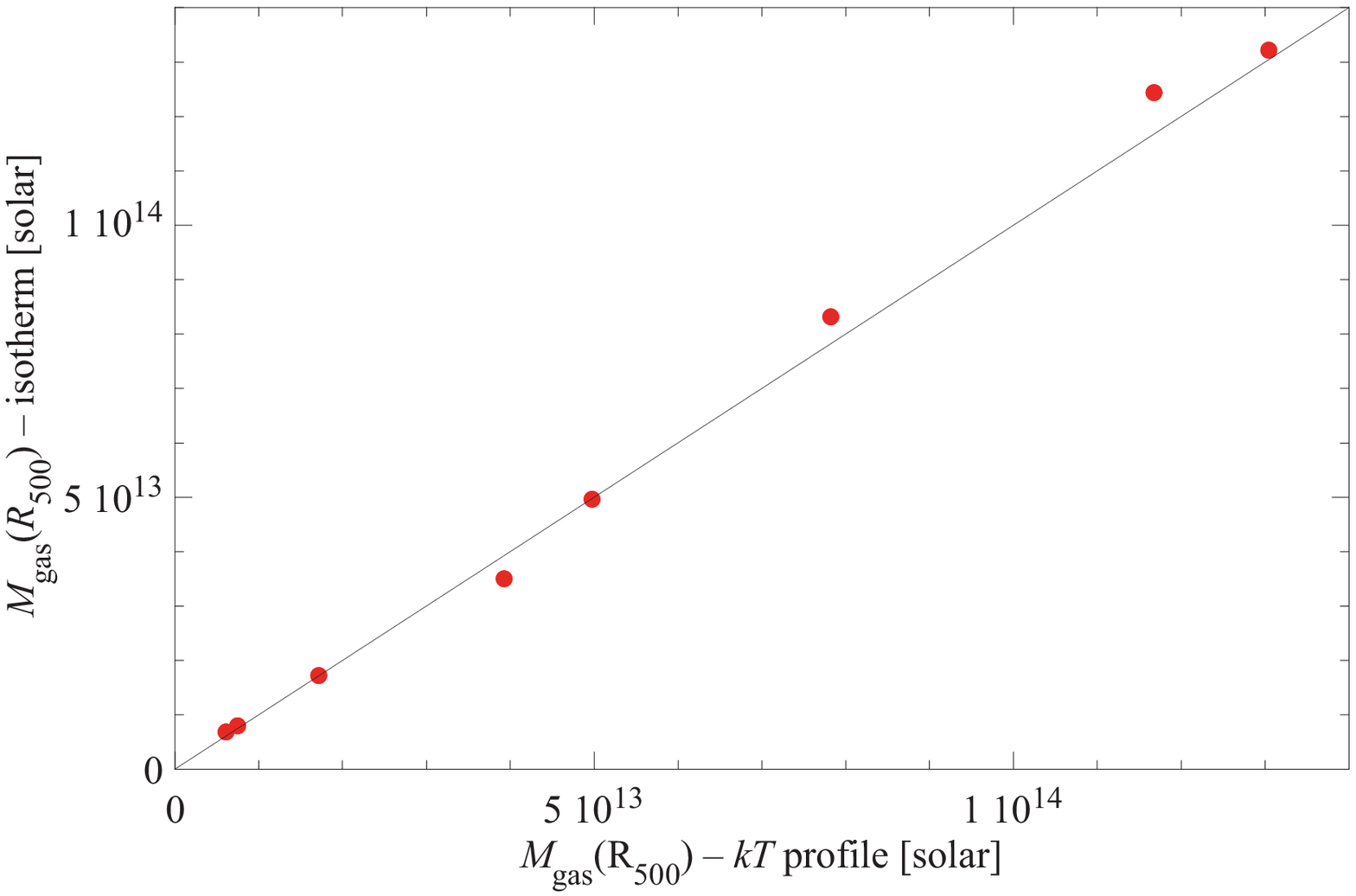}
\caption{\textit{Upper panel}: $r_{500}$ computed assuming an isothermal gas, 
as a function of $r_{500}$ computed, assuming the temperature profile given in \citet{V06}.
\textit{Middle panel}: $M_{\rm tot, 500}$ computed, assuming an isothermal gas, 
as a function of $M_{\rm tot, 500}$ computed, assuming the temperature profile given in \citet{V06}.
\textit{Lower panel}:$M_{\rm gas, 500}$ computed, assuming an isothermal gas as a function 
of $M_{\rm gas, 500}$ computed, assuming the temperature profile given in \citet{V06}.} 
\label{fig:isotherm}
\end{figure}

As we can see from Fig. A.1, the assumption of isothermality does not
introduce any systematic error in either the total mass or  $r_{500}$. Moreover,
the values are in good agreement, which is shown even for cool-core clusters, that the derived
values in this work are robust.

\section{Geometric correction}

The stellar mass is measured in a cone, along the line-of-sight
  of radius $r_{2500}$ or $r_{500}$ at the group/cluster distance. For
  simplicity, we will approximate the cone by a cylinder, since the distance of the cluster 
  is much larger than its radius. The length
  of the cylinder can only be roughly estimated by the given scatter in
  redshift space, when enough galaxies have a redshift determination.
  On the other hand, the gas and dynamical masses are measured in
  spheres of radius $r_{2500}$ or $r_{500}$.  Compared to the sphere,
  the cylinder will cover a greater volume in space and therefore, a
  geometric correction is needed.

To estimate this correction, one must assume (or
  determine) the spatial distribution of galaxies in clusters and
  groups. Using SDSS data, \citet{hansen05} showed that the radial
  profile of the galaxy distribution is well represented by a NFW
  profile \citep{NFW97}, which is shallower than the dark matter distribution
  in clusters with a concentration parameter $c \approx 2$--4.

Given a radial profile, the mass excess,$\Upsilon$, of a cylinder compared to a sphere is:
\begin{equation}
\Upsilon = M_{\rm cyl} / M_{\rm sph} \, ,
\end{equation}
where
\begin{equation}
M_{\rm sph} = 4\pi \int_0^{R_{\rm max}} \rho(r) r^2 d r \quad \mbox{and} \quad
\end{equation}
\begin{equation}
M_{\rm cyl} = 4\pi \int_0^{z_{\rm max}}\!\int_0^{R_{\rm max}} \rho(R,z) R d R \, d z\, .
\end{equation}

Here, $R_{\rm max}$ is the sphere radius of either $r_{500}$ or $r_{2500}$, and $z_{max}$ is 
half the length of the cylinder (i.e., we are measuring the mass 
between $\pm z_{\rm max}$). Although cumbersome, $\Upsilon$ can be determined analytically, 
if we assume a NFW profile. In terms of the 
concentration parameter, $c \equiv r_{200}/r_s$, we have:

\begin{equation}
\begin{split}
 \Upsilon = & \sqrt{a^2 c^2 -1}\times [(1 + a c) \ln(1 + a c) - a c]  \times 
(1+ a c)^{-1} \times \\
 &  \bigg[ \arctan \left( \frac{b c}{\sqrt{a^2 c^2 -1}} \right) - 
\arctan \left(\frac{b}{\sqrt{(a^2+b^2)(a^2 c^2 - 1) }} \right)+ \\
 & \sqrt{a^2 c^2 - 1} \times \ln {\left( \frac{a + a b c}{b + \sqrt{a^2 + b^2}} \right)} \bigg]^{-1}
 % \ln \left( \frac{a + a b c}{b + \sqrt{a^2 + b^2}} \right) 
\, ,
\end{split}
\end{equation}
where $a = R_{\rm max}/r_{200}$ and $b= z_{\rm max}/r_{200}$.

For a NFW profile, $r_{500}/r_{200} = 0.61$--0.65 for $c=2$--4 and  
$r_{2500}/r_{200} = 0.22$--0.28 also for $c=2$--4. The mass excess in the cylinder is shown in Fig. B.1.:

\begin{figure}[htb]
\includegraphics*[width=8.6cm]{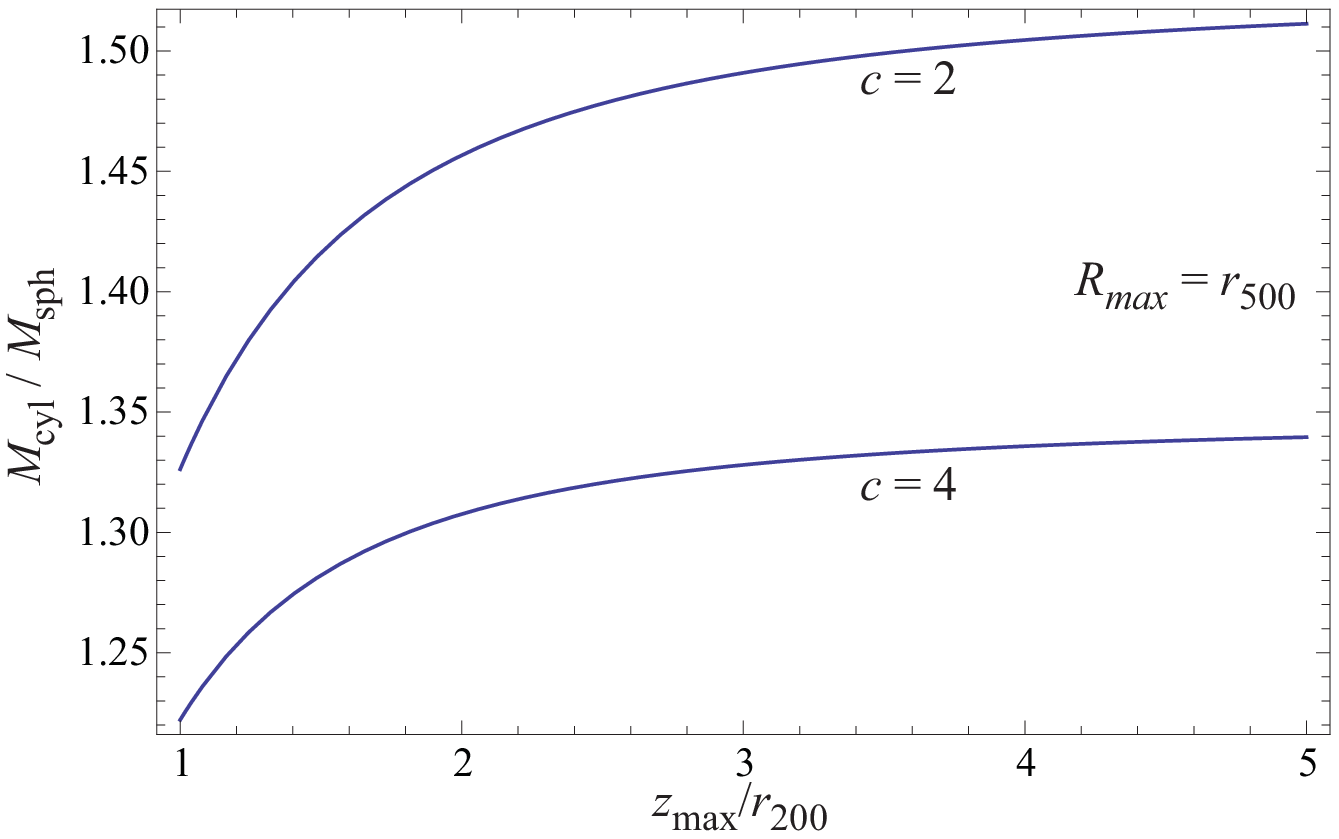}
\includegraphics*[width=8.6cm]{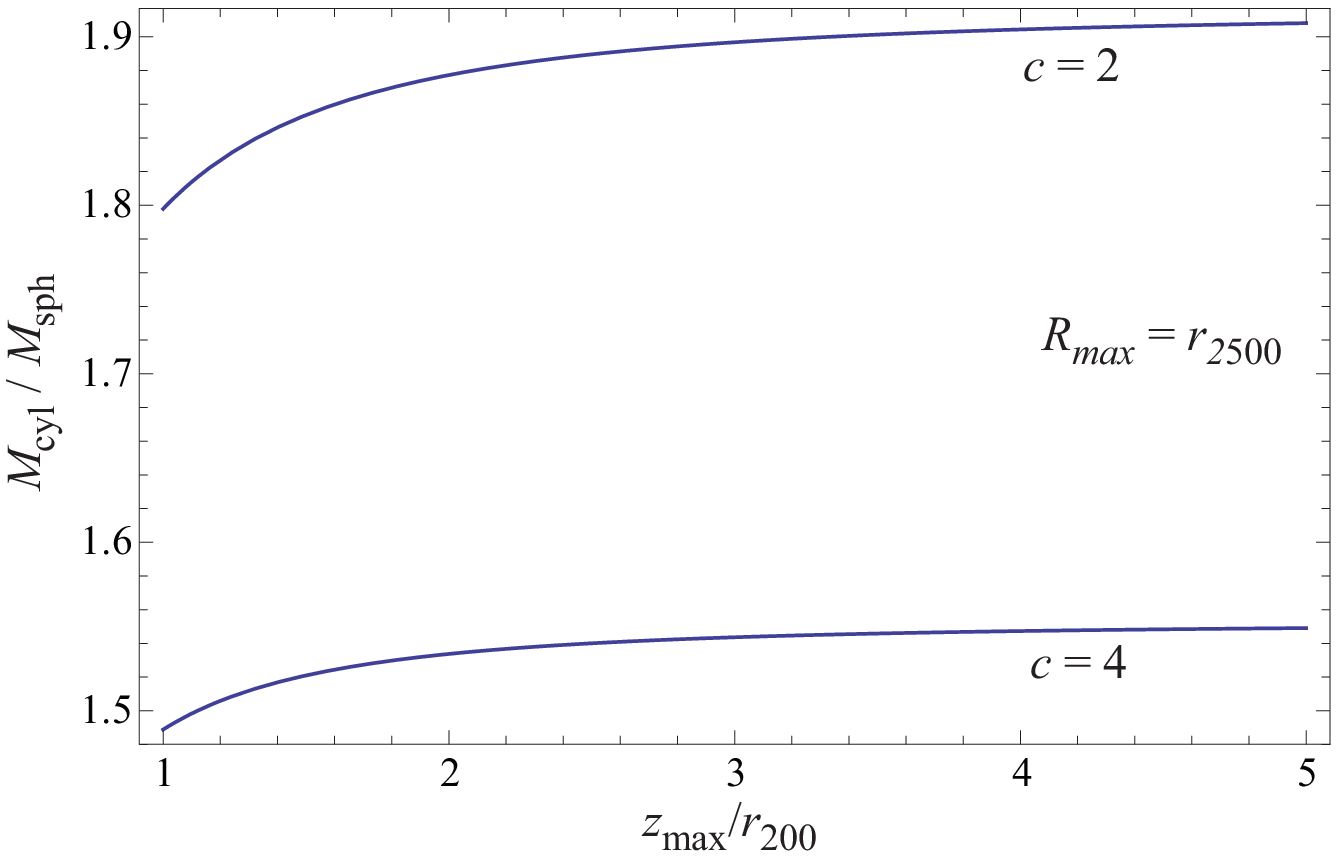}
\caption[]{Upper panel: for a fixed $R_{\rm max} = r_{500}$, mass excess as a function 
of $z_{\rm max}/r_{200}$,  for two NFW
  concentration parameters $c$. Lower panel: same as upper panel but for $R_{\rm
    max} = r_{2500}$.}
\end{figure}

If we assume that we are including galaxies up to $\sim 3 r_{200}$, 
then we very roughly  have the following mass excesses: $\Upsilon \sim 1.4$ 
for spheres of $r_{500}$, and $\Upsilon \sim 1.7$ for spheres of $r_{2500}$.  
To compare the stellar mass in galaxies with the gas and 
dynamical masses, the former should therefore be divided by the above $\Upsilon$, depending on the 
extraction radius.

\end{appendix}

\onecolumn
\begin{landscape}
%\begin{center}
\scriptsize
%\longtab{1}{
\setlength{\tabcolsep}{5pt}
\begin{longtable}{lcccrrrrrrrrr}
\caption{\label{tab:geral} General properties, X-ray parameters and stellar masses derived for our sample}\\
\hline\hline
\multicolumn{13}{c}{Groups}\\
Name & z & RA & Dec &$r_{2500}$ &$r_{500}$& $ <kT> $ & $M_{\rm gas,2500} $ &  $M_{\rm gas,500} $ &  $M_{\rm tot,2500} $ &  $M_{\rm tot,500} $ &    $M_{\rm star,2500} $ &  $M_{\rm star,500} $\\
 & & & & ($h_{70}^{-1}$) kpc & ($h_{70}^{-1}$) kpc &(keV) & $(10^{12} M_{\odot})$ & $(10^{12} M_{\odot})$ & $(10^{13} M_{\odot})$ & $(10^{13} M_{\odot})$ & $(10^{12} M_{\odot})$ & $(10^{12} M_{\odot})$\\
 \hline 
NGC1132   	& 0.023 & 02h52m51.9s & --01d16m29s& 171.5 &383.5 & 0.98 $\pm$ 0.10 & 0.29 $\pm$ 0.03 &1.24 $\pm$ 0.03& 0.85 $\pm$ 0.09 &1.55 $\pm$ 0.17&  0.37$\pm$  0.09& 0.60 $\pm$ 0.12 \\
RBS461    	& 0.030 & 03h41m21.7s & +15d24m18s& 262.4 &586.8 & 2.14 $\pm$ 0.46 & 0.55 $\pm$ 0.09 &2.09 $\pm$ 0.39& 0.92 $\pm$ 0.25 &2.70 $\pm$ 0.73&  0.54  $\pm$  0.15 & 0.95 $\pm$ 0.25 \\
NGC4104   	& 0.028 & 12h06m39.0s & +28d10m27s& 224.7 &502.5 & 1.63 $\pm$ 0.18 & 0.57 $\pm$ 0.05 &2.05 $\pm$ 0.09& 1.16 $\pm$ 0.13 &2.70 $\pm$ 0.48&  0.28 $\pm$  0.07 & 0.69 $\pm$0.18  \\
NGC4325   	& 0.026 & 12h23m06.7s & +10d37m16s& 182.1 &408.6 & 0.89 $\pm$ 0.11 & 0.47 $\pm$ 0.05 &1.50 $\pm$ 0.02& 0.96 $\pm$ 0.12 &2.27 $\pm$ 0.27&  0.14 $\pm$  0.04 & 0.29 $\pm$ 0.08 \\
NGC5098   	& 0.037 & 13h20m16.2s & +33d08m39s& 154.1 &344.6 & 0.94 $\pm$ 0.17 & 0.32 $\pm$ 0.05 &1.30 $\pm$ 0.21& 0.63 $\pm$ 0.11 &1.41 $\pm$ 0.25&  0.31 $\pm$  0.08 & 0.55 $\pm$ 0.14 \\
Abell1991		& 0.059 & 14h54m30.2s & +18d37m51s& 251.9 &563.3 & 2.20 $\pm$ 0.57 & 2.11 $\pm$ 0.46 &6.68 $\pm$ 0.12& 2.85 $\pm$ 0.74 &6.33 $\pm$ 1.66&  0.59 $\pm$  0.18 & 1.09 $\pm$ 0.30 \\
AWM4 	  	& 0.032 & 16h04m57.0s & +23d55m14s& 245.0 &547.8 & 2.40 $\pm$ 0.26 & 1.27 $\pm$ 0.12 &5.60 $\pm$ 0.29& 2.39 $\pm$ 0.26 &5.55 $\pm$ 0.61&  0.42 $\pm$  0.10 & 0.62 $\pm$ 0.14 \\
IC1262 	  	& 0.033 & 17h33m02.0s & +43d45m35s& 284.8 &636.9 & 2.64 $\pm$ 1.51 & 1.75 $\pm$ 0.64 &6.36 $\pm$ 0.65& 3.90 $\pm$ 1.23 &7.32 $\pm$ 4.20&  0.37 $\pm$  0.11 & 0.65 $\pm$ 0.18 \\
RXCJ2315.7-0222 & 0.027 & 23h15m44.1s & --02d22m59s& 191.0 &427.1 & 1.39 $\pm$ 0.23 & 0.51 $\pm$ 0.06 &1.86 $\pm$ 0.19& 1.15 $\pm$ 0.25 &2.37 $\pm$ 0.51&  0.31 $\pm$ 0.09 & 0.70 $\pm$ 0.20\\
\hline
\endfirsthead
\caption{continued.}\\
\hline\hline
Name & z & R.A. & Dec  & $r_{2500}$ & $r_{500}$& $ <kT> $ & $M_{\rm gas,2500} $ &  $M_{\rm gas,500} $ &  $M_{\rm tot,2500} $ &  $M_{\rm tot,500} $&    $M_{\rm star,2500} $ &  $M_{\rm star,500} $\\
& & & &  ($h_{70}^{-1}$) kpc & ($h_{70}^{-1}$) kpc &(keV) & $(10^{13} M_{\odot})$ & $(10^{13} M_{\odot})$ & $(10^{14} M_{\odot})$ & $(10^{14} M_{\odot})$ & $(10^{13} M_{\odot})$ & $(10^{13} M_{\odot})$\\
\hline\hline
\endhead
\hline
\endfoot
\hline
\multicolumn{13}{c}{Clusters from M12}\\
Name & z & RA & Dec &$r_{2500}$ &$r_{500}$& $ <kT> $ & $M_{\rm gas,2500} $ &  $M_{\rm gas,500} $ &  $M_{\rm tot,2500} $ &  $M_{\rm tot,500} $ &    $M_{\rm star,2500} $ &  $M_{\rm star,500} $\\
 & & & & ($h_{70}^{-1}$) kpc & ($h_{70}^{-1}$) kpc &(keV) & $(10^{13} M_{\odot})$ & $(10^{13} M_{\odot})$ & $(10^{14} M_{\odot})$ & $(10^{14} M_{\odot})$ & $(10^{13} M_{\odot})$ & $(10^{13} M_{\odot})$\\
\hline
MS0015.9+1609     &0.54 & 00h18m33.631s &+16d26m13.56s  &  445.6   &  1092.6  &  8.3  $\pm$0.5 & 3.34 $\pm$ 0.17&12.9 $\pm$0.6  & 2.26$\pm$0.14& 6.67 $\pm$0.40& - & -  \\
RXJ0027.6+2616    &0.37 & 00h27m46.051s &+26d16m19.92s  &  328.4   &  914.8   &  4.8  $\pm$1.0 & 0.78 $\pm$ 0.14&4.5  $\pm$0.8  & 0.74$\pm$0.15& 3.19 $\pm$0.66& - & -   \\
CLJ0030+2618	&0.50 & 00h30m33.948s &+26d18m08.28s  &  291.2   &  811.4   &  4.1  $\pm$1.7 & 0.60 $\pm$ 0.21&3.1  $\pm$1.1  & 0.60$\pm$0.25& 2.60 $\pm$1.08& - & -   \\
MS0015.9+1609     &0.54 & 00h18m33.631s &+16d26m13.56s  &  445.6   &  1092.6  &  8.3  $\pm$0.5 & 3.34 $\pm$ 0.17&12.9 $\pm$0.6  & 2.26$\pm$0.14& 6.67 $\pm$0.40& - & -  \\
RXJ0027.6+2616    &0.37 & 00h27m46.051s &+26d16m19.92s  &  328.4   &  914.8   &  4.8  $\pm$1.0 & 0.78 $\pm$ 0.14&4.5  $\pm$0.8  & 0.74$\pm$0.15& 3.19 $\pm$0.66& - & -   \\
CLJ0030+2618	 &0.50 & 00h30m33.948s &+26d18m08.28s  &  291.2   &  811.4   &  4.1  $\pm$1.7 & 0.60 $\pm$ 0.21&3.1  $\pm$1.1  & 0.60$\pm$0.25& 2.60 $\pm$1.08& - & -   \\
A68	 		 &0.26 & 00h37m06.089s &+09d09m33.04s  &  559.6   &  1352.7  &  7.8  $\pm$1.0 & 2.98 $\pm$ 0.32&8.8  $\pm$0.9  & 3.22$\pm$0.41& 9.10 $\pm$1.17& - & -   \\
A115	 		 &0.20 & 00h55m50.688s &+26d24m37.80s  &  429.4   &  1035.7  &  6.7  $\pm$1.0 & 1.45 $\pm$ 0.18&6.5  $\pm$0.8  & 1.37$\pm$0.20& 3.84 $\pm$0.57& 0.14 $\pm$ 0.04&  0.81$\pm$ 0.24\\
A209	 	 	 &0.21 & 01h31m53.520s &--13d36m46.44s  &  479.9   &  1213.1  &  7.4  $\pm$0.5 & 2.23 $\pm$ 0.13&9.7  $\pm$0.5  & 1.93$\pm$0.13& 6.22 $\pm$0.42& - & -   \\
CLJ0152.7-1357S  &0.83 & 01h52m39.888s &--13d58m26.76s  &  239.7   &  625.3     &  4.9  $\pm$1.1 & 0.50 $\pm$ 0.09&3.0  $\pm$0.6  & 0.50$\pm$0.11& 1.77 $\pm$0.40& - 			& -   \\
A267	 		 &0.23 & 01h52m42.144s &+01d00m41.29s  &  404.3   &  953.3   &  4.4  $\pm$0.5 & 1.61 $\pm$ 0.15&5.2  $\pm$0.5  & 1.18$\pm$0.13& 3.10 $\pm$0.35&  0.25 $\pm$ 0.07 & 0.55 $\pm$ 0.16 \\
CLJ0152.7-1357N 	 &0.83  & 01h52m44.280s &--13d57m19.08s   &  239.0   &  621.1   &  4.9  $\pm$0.9 & 0.53 $\pm$ 0.08&3.2  $\pm$0.5  & 0.49$\pm$0.09& 1.73 $\pm$0.32&			-	 &  - \\
MACSJ0159.8-0849&0.41 & 01h59m49.368s &--08d50m00.42s  &  570.7   &  1296.3  &  10.2 $\pm$0.9 & 4.09 $\pm$ 0.30&11.5 $\pm$0.8  & 4.05$\pm$0.36& 9.49 $\pm$0.84& - & -   \\
CLJ0216-1747	 &0.58 & 02h16m32.856s &--17d47m32.28s  &  318.9   &  849.9    &  5.6  $\pm$3.8  & 0.47 $\pm$ 0.27 &2.4    $\pm$1.4  & 0.87$\pm$0.59& 3.28 $\pm$2.22& - & -   \\
RXJ0232.2-4420     &0.28 & 02h32m18.216s &--44d20m49.56s  &  539.0   &  1317.5  &  8.0  $\pm$1.4  & 3.58 $\pm$ 0.52 &11.2 $\pm$1.6  & 2.97$\pm$0.52& 8.68 $\pm$1.52& - & -   \\
MACSJ0242.5-2132 &0.31 & 02h42m35.856s &--21d32m26.16s  &  439.9   &  985.5   &  5.5  $\pm$0.7 & 2.14 $\pm$ 0.23&5.7  $\pm$0.6  & 1.67$\pm$0.21& 3.76 $\pm$0.48& - & -   \\
A383	 	 & 0.19 & 02h48m03.432s &--03d31m45.87s  &  425.1   &  971.2   &  4.5  $\pm$0.3 & 1.40 $\pm$ 0.08&4.0  $\pm$0.2  & 1.31$\pm$0.09& 3.13 $\pm$0.21& 0.22 $\pm$ 0.07& 0.49 $\pm$ 0.15  \\
MACSJ0257.6-2209 & 0.32 & 02h57m41.328s &--22d09m14.40s  &  455.5   &  1199.4  &  6.7  $\pm$0.9 & 1.99 $\pm$ 0.22&7.6  $\pm$0.9  & 1.87$\pm$0.25& 6.83 $\pm$0.92& - & -   \\
MS0302.7+1658   &0.42 & 03h05m31.656s &+17d10m10.56s  &  308.9   &  697.8   &  3.3  $\pm$0.8 & 0.82 $\pm$ 0.17&2.5  $\pm$0.5  & 0.66$\pm$0.16& 1.51 $\pm$0.37& - & -   \\
CLJ0318-0302	 &0.37 & 03h18m33.456s &--03d02m57.30s  &  435.4   &  1030.8  &  5.4  $\pm$1.7 & 1.31 $\pm$ 0.34&3.8  $\pm$1.0  & 1.73$\pm$0.54& 4.58 $\pm$1.44& - & -   \\
MACSJ0329.6-0211 &0.45 & 03h29m41.520s &--02d11m45.99s  &  349.9   &  839.6   &  4.5  $\pm$0.5 & 1.74 $\pm$ 0.16&5.7  $\pm$0.5  & 0.98$\pm$0.11& 2.72 $\pm$0.30& - & -   \\
MACSJ0404.6+1109&0.36 & 04h04m32.664s &+11d08m16.80s  &  343.1   &  834.5   &  5.5  $\pm$0.7 & 0.84 $\pm$ 0.09&4.5  $\pm$0.5  & 0.83$\pm$0.11& 2.39 $\pm$0.30& - & -   \\
MACSJ0429.6-0253&0.40 & 04h29m35.952s &--02d53m06.21s  &  456.1   &  1065.8  &  6.5  $\pm$0.7 & 2.27 $\pm$ 0.20&6.5  $\pm$0.6  & 2.05$\pm$0.22& 5.24 $\pm$0.56& - & -   \\
RXJ0439.0+0715  &0.23 & 04h39m00.672s &+07d16m03.75s  &  440.1   &  1169.6  &  5.3  $\pm$0.4 & 2.05 $\pm$ 0.13&7.3  $\pm$0.5  & 1.52$\pm$0.12& 5.72 $\pm$0.43& - & -   \\
RXJ0439+0520	 &0.21 & 04h39m02.304s &+05d20m43.26s  &  387.7   &  848.8   &  3.9  $\pm$0.4 & 1.15 $\pm$ 0.10&3.1  $\pm$0.3  & 1.02$\pm$0.10& 2.14 $\pm$0.22& - & -   \\
MACSJ0451.9+0006 &0.43 & 04h51m54.408s &+00d06m19.13s  &  363.0   &  892.7   &  5.0  $\pm$1.1 & 1.50 $\pm$ 0.27&5.9  $\pm$1.1  & 1.07$\pm$0.24& 3.19 $\pm$0.70& - & -   \\
A521	 	 &0.55 & 04h54m11.184s &--03d00m53.85s  &  292.4   &  906.2   &  4.8  $\pm$0.2 & 0.67 $\pm$ 0.02&6.8  $\pm$0.2  & 0.46$\pm$0.02& 2.73 $\pm$0.11& - & -   \\
A520	 	 &0.25 & 04h54m06.576s &--10d13m15.24s  &  447.1   &  1251.2  &  6.5  $\pm$0.3 & 2.02 $\pm$ 0.08&10.3 $\pm$0.4  & 1.55$\pm$0.07& 6.78 $\pm$0.31& - & -   \\
MS0451.6-0305   &0.20 & 04h54m09.600s &+02d55m28.63s  &  447.1   &  1148.9  &  7.6  $\pm$1.2 & 3.60 $\pm$ 0.47&12.6 $\pm$1.7  & 2.31$\pm$0.36& 7.83 $\pm$1.24& - & -   \\
CLJ0522-3625	 &0.47 & 05h22m14.832s &--36d24m57.60s  &  318.7   &  845.2   &  4.3  $\pm$1.4 & 0.60 $\pm$ 0.16&2.8  $\pm$0.7  & 0.76$\pm$0.25& 2.84 $\pm$0.93& - & -   \\
CLJ0542.8-4100  &0.64 & 05h42m50.208s &--41d00m04.32s  &  346.4   &  872.4   &  6.2  $\pm$1.0 & 1.11 $\pm$ 0.15&4.8  $\pm$0.7  & 1.20$\pm$0.19& 3.83 $\pm$0.62& - & -   \\
MACSJ0647.7+7015 &0.58 & 06h47m50.160s &+70d14m55.68s  &  533.6   &  1231.0  &  11.3 $\pm$2.1 & 3.88 $\pm$ 0.60&11.8 $\pm$1.8  & 4.09$\pm$0.76& 10.04$\pm$1.86& - & -   \\
1E0657-56	 &0.30 & 06h58m30.000s &--55d56m39.12s  &  692.9   &  1722.6  &  11.7 $\pm$0.5 & 8.57 $\pm$ 0.31&25.0 $\pm$0.9  & 6.40$\pm$0.27& 19.66$\pm$0.84& - & -   \\
MACSJ0717.5+3745 &0.55 & 07h17m31.680s &+37d45m31.68s  &  400.9   &  1513.0  &  10.6 $\pm$1.0 & 2.77 $\pm$ 0.22&24.4 $\pm$1.9  & 1.66$\pm$0.16& 17.81$\pm$1.68& - & -   \\
A586	 	 &0.17 & 07h32m20.160s &+31d37m55.92s  &  557.8   &  1265.8  &  7.6  $\pm$0.8 & 2.45 $\pm$ 0.22&6.9  $\pm$0.6  & 2.91$\pm$0.31& 6.81 $\pm$0.72& 0.2 $\pm$ 0.06& 2.97 $\pm$ 0.89 \\
MACSJ0744.9+3927&0.70 & 07h44m52.320s &+39d27m26.64s  &  392.2   &  1017.6  &  8.1  $\pm$0.7 & 2.60 $\pm$ 0.19&10.3 $\pm$0.7  & 1.86$\pm$0.16& 6.49 $\pm$0.56& - & -   \\
A665		 &0.18 & 08h30m57.360s &+65d50m33.36s  &  481.6   &  1357.3  &  7.8  $\pm$0.4 & 2.39 $\pm$ 0.10&12.6 $\pm$0.5  & 1.90$\pm$0.10& 8.50 $\pm$0.44& 0.43 $\pm$ 0.13 & 0.58 $\pm$ 0.17 \\
A697		 &0.28 & 08h42m57.600s &+36d21m55.80s  &  578.7   &  1455.7  &  10.2 $\pm$0.8 & 4.34 $\pm$ 0.28&16.1 $\pm$1.1  & 3.67$\pm$0.29& 11.68$\pm$0.92& 0.34 $\pm$ 0.11 &  0.70 $\pm$ 0.21\\
CLJ0848.7+4456  &0.57 & 08h48m47.760s &+44d56m13.92s  &  199.7   &  460.7   &  2.0  $\pm$0.2 & 0.19 $\pm$ 0.02&0.8  $\pm$0.1  & 0.21$\pm$0.02& 0.52 $\pm$0.05& - & -   \\
ZWCLJ1953	 &0.32 & 08h50m06.960s &+36d04m17.40s  &  441.9   &  1057.0  &  6.1  $\pm$0.6 & 1.96 $\pm$ 0.16&6.8  $\pm$0.6  & 1.71$\pm$0.17& 4.67 $\pm$0.46& - & -   \\
CLJ0853+5759	 &0.48 & 08h53m14.880s &+57d59m57.48s  &  298.5   &  943.1   &  5.1  $\pm$1.5 & 0.42 $\pm$ 0.10&3.1  $\pm$0.8  & 0.63$\pm$0.18& 3.97 $\pm$1.17& - & -   \\
MS0906.5+1110   &0.18 & 09h09m12.720s &+10d58m32.88s  &  405.7   &  923.0   &  4.7  $\pm$0.3 & 1.28 $\pm$ 0.07&4.4  $\pm$0.2  & 1.13$\pm$0.07& 2.67 $\pm$0.17& 0.28 $\pm$0.09 & 0.92 $\pm$0.31  \\
RXJ0910+5422	 &1.11 & 09h10m44.880s &+54d22m07.68s  &  161.6   &  524.9   &  2.7  $\pm$1.9 & 0.20 $\pm$ 0.12&1.3  $\pm$0.8  & 0.21$\pm$0.15& 1.45 $\pm$1.02& - & -   \\
A773	 	 &0.22 & 09h17m53.040s &+51d43m39.72s  &  506.2   &  1306.3  &  7.4  $\pm$0.4 & 2.52 $\pm$ 0.11&9.6  $\pm$0.4  & 2.29$\pm$0.12& 7.86 $\pm$0.42& 0.20 $\pm$0.06& 0.80 $\pm$0.24  \\
A781		 &0.30 & 09h20m26.160s &+30d30m02.52s  &  332.0   &  984.4   &  5.5  $\pm$0.5 & 0.85 $\pm$ 0.06&6.6  $\pm$0.5  & 0.71$\pm$0.06& 3.68 $\pm$0.33& 0.30 $\pm$0.10 & 0.60 $\pm$0.18  \\
CLJ0926+1242	 &0.49 & 09h26m36.480s &+12d43m03.36s  &  306.2   &  804.2   &  4.5  $\pm$1.0 & 0.61 $\pm$ 0.11&2.9  $\pm$0.5  & 0.69$\pm$0.15& 2.50 $\pm$0.56& - & -   \\
RBS797 	 &0.35 & 09h47m12.960s &+76d23m13.56s  &  509.5   &  1139.7  &  7.3  $\pm$1.0 & 3.30 $\pm$ 0.38&8.1  $\pm$0.9  & 2.72$\pm$0.37& 6.08 $\pm$0.83& - & -   \\
MACSJ0949.8+1708 &0.38 & 09h49m51.840s &+17d07m07.68s  &  457.2   &  1057.7  &  7.3  $\pm$0.9 & 2.59 $\pm$ 0.27&9.0  $\pm$0.9  & 2.03$\pm$0.25& 5.03 $\pm$0.62& - & -   \\
CLJ0956+4107	 &0.59 & 09h56m03.360s &+41d07m12.72s  &  278.8   &  792.9   &  4.0  $\pm$0.7 & 0.61 $\pm$ 0.09&3.2  $\pm$0.5  & 0.58$\pm$0.10& 2.69 $\pm$0.47& - & -   \\
A907		 &0.15 & 09h58m21.840s &--11d03m49.32s  &  471.0   &  1063.7  &  5.4  $\pm$0.2 & 1.79 $\pm$ 0.06&5.2  $\pm$0.2  & 1.72$\pm$0.06& 3.97 $\pm$0.15& - & -   \\
MS1006.0+1202   &0.22 & 10h08m47.520s &+11d47m42.36s  &  469.4   &  1122.1  &  6.0  $\pm$0.5 & 1.58 $\pm$ 0.11&5.4  $\pm$0.4  & 1.83$\pm$0.15& 5.01 $\pm$0.42& - & -   \\
MS1008.1-1224   &0.30 & 10h10m32.400s &--12d39m30.24s  &  353.4   &  997.7   &  4.6  $\pm$0.4 & 1.13 $\pm$ 0.08&5.5  $\pm$0.4  & 0.85$\pm$0.07& 3.84 $\pm$0.33& - & -   \\
ZW3146 	 &0.29 & 10h23m39.600s &+04d11m12.08s  &  561.6   &  1267.8  &  7.8  $\pm$0.4 & 4.18 $\pm$ 0.18&10.1 $\pm$0.4  & 3.39$\pm$0.17& 7.79 $\pm$0.40& - & -   \\
CLJ1113.1-2615  &0.73 & 11h13m05.280s &-26d15m38.88s  &  262.2   &  610.7   &  3.6  $\pm$1.3 & 0.50 $\pm$ 0.15&1.8  $\pm$0.5  & 0.57$\pm$0.21& 1.45 $\pm$0.52& - & -   \\
A1204   	 &0.17 &11h13m20.400s  & +17d35m40.92s &  383.0   &  858.1   &  3.7  $\pm$0.3 & 1.10 $\pm$ 0.07&3.0  $\pm$0.2  & 0.94$\pm$0.08& 2.12 $\pm$0.17& 0.29 $\pm$0.09 & 0.19 $\pm$0.06  \\
CLJ1117+1745	 &0.55 & 11h17m30.000s &+17d44m52.44s  &  245.6   &  569.4   &  3.3  $\pm$0.9 & 0.29 $\pm$ 0.07&1.3  $\pm$0.3  & 0.38$\pm$0.10& 0.95 $\pm$0.26& - & -   \\
CLJ1120+4318	 &0.56 & 11h20m57.360s &+23d26m30.12s  &  341.2   &  906.2   &  5.2  $\pm$1.3 & 1.33 $\pm$ 0.28&5.3  $\pm$1.1  & 1.09$\pm$0.27& 4.08 $\pm$1.02& - & -   \\
RXJ1121+2327	 &0.60 & 11h20m07.200s &+43d18m05.76s  &  51.5    &  709.7   &  3.2  $\pm$0.3 & 0.00 $\pm$ 0.00&3.0  $\pm$0.2  & 0.00$\pm$0.00& 1.87 $\pm$0.18& - & -   \\
A1240   	 &0.16 & 11h23m37.920s &+43d05m37.32s  &  111.2   &  1108.8  &  3.8  $\pm$0.3 & 0.01 $\pm$ 0.00&3.6  $\pm$0.2  & 0.02$\pm$0.00& 4.52 $\pm$0.36& - & -   \\
MACSJ1131.8-1955 &0.31 & 11h31m55.200s &--19d55m50.88s  &  501.0   &  1466.2  &  9.5  $\pm$1.8 & 2.88 $\pm$ 0.45&15.5 $\pm$2.4  & 2.45$\pm$0.46& 12.27$\pm$2.33& - & -   \\
MS1137.5+6625   &0.78 & 11h40m22.320s &+66d08m15.72s  &  368.7   &  853.6   &  6.5  $\pm$1.4 & 1.46 $\pm$ 0.26&4.2  $\pm$0.8  & 1.71$\pm$0.37& 4.24 $\pm$0.91& - & -   \\
MACSJ1149.5+2223&0.55 & 11h49m35.040s &+22d24m10.08s  &  402.8   &  1044.2  &  8.5  $\pm$1.1 & 2.32 $\pm$ 0.25&12.2 $\pm$1.3  & 1.68$\pm$0.22& 5.85 $\pm$0.76& - & -   \\
A1413   	 &0.14 & 11h55m18.000s &+23d24m17.28s  &  556.4   &  1330.7  &  7.1  $\pm$0.3 & 2.84 $\pm$ 0.10&8.4  $\pm$0.3  & 2.81$\pm$0.12& 7.69 $\pm$0.32&  0.42  $\pm$ 0.13 & 0.74 $\pm$0.22  \\
CLJ1213+0253	 &0.41 & 12h13m34.800s &+02d53m46.39s  &  316.0   &  809.7   &  3.9  $\pm$0.9 & 0.49 $\pm$ 0.09&2.1  $\pm$0.4  & 0.69$\pm$0.16& 2.32 $\pm$0.54& - & -   \\
RXJ1221+4918	 &0.70 & 12h21m26.400s &+49d18m30.24s  &  302.2   &  789.3   &  5.9  $\pm$0.7 & 0.85 $\pm$ 0.08&4.6  $\pm$0.5  & 0.85$\pm$0.10& 3.04 $\pm$0.36& - & -   \\
CLJ1226.9+3332  &0.89 & 12h26m57.840s &+33d32m47.76s  &  436.8   &  1029.6  &  10.0 $\pm$1.9 & 3.08 $\pm$ 0.49&8.7  $\pm$1.4  & 3.23$\pm$0.61& 8.46 $\pm$1.61& - & -   \\
RXJ1234.2+0947  &0.23 & 12h34m17.50s &+09d45m58.48s  &  382.0   &  1477.9  &  7.6  $\pm$2.4 & 0.73 $\pm$ 0.19&8.3  $\pm$2.2  & 1.00$\pm$0.31& 11.53$\pm$3.64& 0.23 $\pm$0.08 & 0.47 $\pm$0.14  \\
RDCS1252-29	 &1.24 & 12h52m54.960s &--29d27m20.88s  &  195.2   &  503.3   &  4.6  $\pm$0.9 & 0.33 $\pm$ 0.05&1.7  $\pm$0.3  & 0.43$\pm$0.08& 1.47 $\pm$0.29& - & -   \\
A1682   	 &0.23 & 13h06m50.880s &+46d33m30.24s  &  388.0   &  939.2   &  5.8  $\pm$2.0 & 1.15 $\pm$ 0.33&5.6  $\pm$1.6  & 1.05$\pm$0.36& 2.98 $\pm$1.03& 0.20 $\pm$0.06 & 0.52 $\pm$0.16  \\
MACSJ1311.0-0310 &0.49 & 13h11m01.680s &--03d10m36.87s  &  450.6   &  1030.4  &  6.5  $\pm$0.8 & 2.01 $\pm$ 0.21&5.1  $\pm$0.5  & 2.21$\pm$0.27& 5.29 $\pm$0.65& - & -   \\
A1689 		 &0.18 & 13h11m29.520s &--01d20m29.68s  &  619.6   &  1487.5  &  8.4  $\pm$0.4 & 4.56 $\pm$ 0.18&12.0 $\pm$0.5  & 4.05$\pm$0.19& 11.20$\pm$0.53& 0.58 $\pm$0.17 & 1.17 $\pm$0.35  \\
RXJ1317.4+2911  &0.81 & 13h17m20.880s &+29d11m15.00s  &  161.2   &  418.9   &  2.2  $\pm$0.8 & 0.11 $\pm$ 0.03&0.7  $\pm$0.2  & 0.15$\pm$0.05& 0.51 $\pm$0.19& - & -   \\
CLJ1334+5031	 &0.62 & 13h34m19.200s &+50d31m05.52s  &  324.4   &  742.2   &  5.2  $\pm$2.1 & 0.89 $\pm$ 0.30&3.3  $\pm$1.1  & 0.96$\pm$0.39& 2.30 $\pm$0.93& - & -   \\
A1763 	 	 &0.22 & 13h35m18.240s &+40d59m59.28s  &  496.4   &  1328.3  &  8.1  $\pm$0.5 & 2.45 $\pm$ 0.13&11.5 $\pm$0.6  & 2.17$\pm$0.13& 8.32 $\pm$0.51& 0.25 $\pm$0.07 & 0.79 $\pm$0.24  \\
RXJ1347.5-1145  &0.45 & 13h47m30.720s &--11d45m10.44s  &  672.0   &  1524.6  &  14.2 $\pm$1.4 & 7.95 $\pm$ 0.65&20.2 $\pm$1.7  & 6.97$\pm$0.69& 16.28$\pm$1.61& - & -   \\
RXJ1350.0+6007  &0.80 & 13h50m48.480s &+60d07m5.520s  &  240.6   &  602.6   &  4.5  $\pm$1.0 & 0.41 $\pm$ 0.08&2.2  $\pm$0.4  & 0.49$\pm$0.11& 1.53 $\pm$0.34& - & -   \\
CLJ1354-0221	 &0.55 & 13h54m17.280s &--02d21m50.97s  &  228.7   &  585.3   &  3.1  $\pm$0.9 & 0.30 $\pm$ 0.07&1.7  $\pm$0.4  & 0.31$\pm$0.09& 1.03 $\pm$0.30& - & -   \\
CLJ1415.1+3612  &1.03 & 14h15m11.040s &+36d12m03.60s  &  229.7   &  574.1   &  4.3  $\pm$0.6 & 0.64 $\pm$ 0.07&2.7  $\pm$0.3  & 0.55$\pm$0.08& 1.73 $\pm$0.24& - & -   \\
RXJ1416+4446	 &0.40 & 14h16m28.080s &+44d46m43.32s  &  321.9   &  793.7   &  3.9  $\pm$0.5 & 0.78 $\pm$ 0.08&3.0  $\pm$0.3  & 0.72$\pm$0.09& 2.17 $\pm$0.28& - & -   \\
MACSJ1423.8+2404&0.54 & 14h23m47.760s &+24d04m41.88s  &  380.1   &  943.4   &  6    $\pm$1.0 & 1.90 $\pm$ 0.26&6.4  $\pm$0.9  & 1.41$\pm$0.23& 4.30 $\pm$0.72& - & -   \\
A1914  	 &0.17 & 14h26m00.960s &+37d49m33.96s  &  623.9   &  1503.5  &  8.5  $\pm$0.6 & 4.36 $\pm$ 0.26&11.7 $\pm$0.7  & 4.08$\pm$0.29& 11.42$\pm$0.81& 0.38 $\pm$0.12 & 0.81 $\pm$0.24 \\
A1942  	 &0.22 & 14h38m22.080s &+03d40m06.38s  &  337.6   &  818.6   &  4.2  $\pm$0.3 & 0.59 $\pm$ 0.04&2.8  $\pm$0.2  & 0.68$\pm$0.05& 1.95 $\pm$0.14& 0.17 $\pm$0.05 & 0.51 $\pm$0.16  \\
MS1455.0+2232   &0.26 & 14h57m15.120s &+22d20m35.52s  &  414.7   &  930.3   &  4.7  $\pm$0.2 & 1.83 $\pm$ 0.06&5.0  $\pm$0.2  & 1.31$\pm$0.06& 2.97 $\pm$0.13& 0.21 $\pm$0.06 & 0.33 $\pm$0.10  \\
RXJ1504-0248	 &0.22 & 15h04m07.440s &--02d48m18.50s  &  621.6   &  1388.6  &  9.4  $\pm$1.1 & 4.40 $\pm$ 0.43&10.7 $\pm$1.0  & 4.23$\pm$0.49& 9.43 $\pm$1.10& 0.21 $\pm$0.08 & 0.38 $\pm$0.12  \\
A2034   	 &0.11 & 15h10m12.480s &+33d30m28.08s  &  499.2   &  1315.5  &  6.3  $\pm$0.2 & 1.85 $\pm$ 0.05&7.3  $\pm$0.2  & 1.97$\pm$0.06& 7.21 $\pm$0.23& 0.07 $\pm$0.02 & 1.71 $\pm$0.50  \\
A2069  	 &0.12 & 15h24m39.840s &+29d53m26.33s  &  334.1   &  1173.2  &  5.9  $\pm$0.3 & 0.51 $\pm$ 0.02&6.3  $\pm$0.3  & 0.59$\pm$0.03& 5.13 $\pm$0.26& 0.27 $\pm$0.09 & 0.22 $\pm$0.06  \\
RXJ1525+0958	 &0.12 & 15h24m09.600s &+29d53m07.80s  &  218.0   &  687.6   &  3.5  $\pm$0.4 & 0.32 $\pm$ 0.03&3.0  $\pm$0.3  & 0.26$\pm$0.03& 1.61 $\pm$0.18& - & -   \\
RXJ1532.9+3021  &0.35 & 15h32m53.760s &+30d20m59.28s  &  476.2   &  1068.4  &  6.3  $\pm$1.0 & 2.89 $\pm$ 0.38&7.2  $\pm$1.0  & 2.20$\pm$0.35& 4.96 $\pm$0.79& - & -   \\
A2111  	 &0.23 & 15h39m41.280s &+34d25m10.92s  &  441.0   &  1035.7  &  6.4  $\pm$0.7 & 1.55 $\pm$ 0.14&6.3  $\pm$0.6  & 1.53$\pm$0.17& 3.97 $\pm$0.43& 0.28 $\pm$0.09 & 0.65 $\pm$0.19 \\
A2125  	 &0.25 & 15h41m08.880s &+66d15m53.28s  &  166.1   &  616.2   &  2.4  $\pm$0.2 & 0.08 $\pm$ 0.01&1.5  $\pm$0.1  & 0.08$\pm$0.01& 0.85 $\pm$0.07& - & -   \\
A2163  	 &0.20 & 16h15m45.840s &--06d08m54.52s  &  722.0   &  2203.3  &  15.2 $\pm$1.2 & 8.45 $\pm$ 0.56&38.7 $\pm$2.5  & 6.54$\pm$0.52& 37.16$\pm$2.93& - & -   \\
MACSJ1621.3+3810 &0.46 & 16h21m24.720s &+38d10m09.48s  &  422.1   &  971.9   &  6.2  $\pm$0.5 & 1.87 $\pm$ 0.13&5.6  $\pm$0.4  & 1.75$\pm$0.14& 4.28 $\pm$0.35& - & -   \\
MS1621.5+2640   &0.43 & 16h23m35.520s &+26d34m20.64s  &  357.7   &  1000.7  &  5.8  $\pm$0.7 & 1.05 $\pm$ 0.11&6.0  $\pm$0.6  & 1.02$\pm$0.12& 4.47 $\pm$0.54& - & -   \\
A2204   	 &0.15 & 16h32m47.040s &+05d34m32.52s  &  586.4   &  1323.8  &  8.4  $\pm$0.8 & 3.78 $\pm$ 0.30&10.6 $\pm$0.8  & 3.32$\pm$0.32& 7.64 $\pm$0.73& - & -   \\
A2218   	 &0.18 & 16h35m52.320s &+66d12m36.00s  &  479.2   &  1130.0  &  6.0  $\pm$0.3 & 2.08 $\pm$ 0.09&7.0  $\pm$0.3  & 1.86$\pm$0.09& 4.87 $\pm$0.24& - & -   \\
CLJ1641+4001	 &0.46 & 16h41m53.040s &+40d01m27.48s  &  291.5   &  742.9   &  3.5  $\pm$0.6 & 0.50 $\pm$ 0.07&2.2  $\pm$0.3  & 0.58$\pm$0.10& 1.91 $\pm$0.33& - & -   \\
RXJ1701+6414	 &0.45 & 17h01m22.800s &+64d14m11.40s  &  311.5   &  723.4   &  4.1  $\pm$0.5 & 0.77 $\pm$ 0.08&3.0  $\pm$0.3  & 0.70$\pm$0.08& 1.74 $\pm$0.21& - & -   \\
RXJ1716.9+6708  &0.81 & 17h16m49.440s &+67d08m25.80s  &  310.4   &  731.5   &  5.7  $\pm$1.1 & 1.05 $\pm$ 0.17&3.9  $\pm$0.6  & 1.06$\pm$0.20& 2.77 $\pm$0.53& - & -   \\
A2259   	 &0.16 & 17h20m10.080s &+27d39m03.28s  &  461.6   &  1117.1  &  5.2  $\pm$0.4 & 1.70 $\pm$ 0.11&5.5  $\pm$0.4  & 1.64$\pm$0.13& 4.65 $\pm$0.36& 0.24 $\pm$0.07 &1.50 $\pm$0.45  \\
RXJ1720.1+2638  &0.39 & 17h20m16.800s &+26d38m06.60s  &  522.0   &  1178.4  &  6.8  $\pm$0.5 & 2.57 $\pm$ 0.16&7.3  $\pm$0.4  & 2.37$\pm$0.17& 5.46 $\pm$0.40& 0.30 $\pm$0.09 & 0.56 $\pm$0.17 \\
MACSJ1720.2+3536 &0.16 & 17h20m08.400s &+27d40m10.20s  &  479.7   &  1176.6  &  7.2  $\pm$0.9 & 2.62 $\pm$ 0.27&8.0  $\pm$0.8  & 2.35$\pm$0.29& 6.95 $\pm$0.87& - & -   \\
A2261  	 &0.22 & 17h22m27.120s &+32d07m56.28s  &  501.7   &  1176.8  &  7.3  $\pm$0.4 & 2.90 $\pm$ 0.13&9.7  $\pm$0.4  & 2.24$\pm$0.12& 5.79 $\pm$0.32 & 0.20 $\pm$0.06 & 0.92 $\pm$0.28 \\
A2294  	 &0.18 & 17h24m09.120s &+85d53m09.96s  &  550.7   &  1588.9  &  8.4  $\pm$1.1 & 2.58 $\pm$ 0.28&10.4 $\pm$1.1  & 2.83$\pm$0.37& 13.57$\pm$1.78& - & -   \\
MACSJ1824.3+4309 &0.49 & 18h24m18.960s &+43d09m48.96s  &  267.0   &  742.4   &  4.8  $\pm$1.4 & 0.39 $\pm$ 0.09&2.9  $\pm$0.7  & 0.46$\pm$0.13& 1.96 $\pm$0.57& - & -   \\
MACSJ1931.8-2634 &0.35 & 19h31m49.680s &--26d34m32.88s  &  472.2   &  1066.4  &  6.7  $\pm$1.1 & 3.13 $\pm$ 0.43&8.6  $\pm$1.2  & 2.16$\pm$0.35& 4.97 $\pm$0.82& - & -   \\
RXJ2011.3-5725  &0.28 & 20h11m27.360s &--57d25m09.84s  &  364.8   &  829.3   &  3.6  $\pm$0.4 & 1.02 $\pm$ 0.09&2.9  $\pm$0.3  & 0.92$\pm$0.10& 2.15 $\pm$0.24& - & -   \\
MS2053.7-0449   &0.58 & 20h56m21.120s &--04d37m47.02s  &  291.3   &  669.0   &  3.9  $\pm$0.6 & 0.62 $\pm$ 0.08&2.2  $\pm$0.3  & 0.66$\pm$0.10& 1.61 $\pm$0.25& - & -   \\
MACSJ2129.4-0741&0.59 & 21h29m25.920s &--07d41m30.26s &  445.6   &  1036.3  &  8.3  $\pm$1.1 & 2.83 $\pm$ 0.31&9.2  $\pm$1.0  & 2.41$\pm$0.32& 6.06 $\pm$0.80& - & -   \\
RXJ2129.6+0005  &0.24 & 21h29m40.080s & +00d05m20.31s &  465.5   &  1264.1  &  6.2  $\pm$0.6 & 2.32 $\pm$ 0.19&8.6  $\pm$0.7  & 1.81$\pm$0.18& 7.26 $\pm$0.70& 0.26 $\pm$0.08 & 0.60 $\pm$0.19  \\
A2409 		  &0.15 & 22h00m52.800s & +20d58m27.84s &  497.5   &  1177.1  &  5.7  $\pm$0.4 & 2.12 $\pm$ 0.12&6.4  $\pm$0.4  & 2.02$\pm$0.14& 5.35 $\pm$0.38& 0.23 $\pm$0.07 & 0.47 $\pm$0.14  \\
MACSJ2228.5+2036 &0.41 & 22h28m32.880s & +20d37m11.64s &  463.4   &  1265.0  &  8.6  $\pm$1.4 & 2.75 $\pm$ 0.37&12.9 $\pm$1.7  & 2.18$\pm$0.36& 8.89 $\pm$1.45& - & -   \\
MACSJ2229.7-2755 &0.32 & 22h29m45.360s & --27d55m36.48s &  402.4   &  1098.4  &  5.0  $\pm$0.9 & 1.69 $\pm$ 0.25&5.9  $\pm$0.9  & 1.29$\pm$0.23& 5.26 $\pm$0.95& - & -   \\
MACSJ2245.0+2637 &0.30 & 22h45m04.800s & +26d38m03.48s &  425.1   &  983.6   &  4.9  $\pm$0.5 & 1.86 $\pm$ 0.16&5.2  $\pm$0.4  & 1.49$\pm$0.15& 3.68 $\pm$0.38& - & -   \\
RXJ2247+0337	 &0.20 & 22h47m28.080s & +03d36m57.78s &  320.2   &  769.9   &  2.9  $\pm$0.9 & 0.22 $\pm$ 0.06&0.8  $\pm$0.2  & 0.57$\pm$0.18& 1.58 $\pm$0.49& - & -   \\
AS1063 	 &0.35 & 22h48m44.880s &--44d31m44.40s &  634.7   &  1453.2  &  11.2 $\pm$1.1 & 7.16 $\pm$ 0.59&19.4 $\pm$1.6  & 5.21$\pm$0.51& 12.52$\pm$1.23& - & -   \\
CLJ2302.8+0844 &0.72 & 23h02m48.000s & +08d43m51.56s &  300.9   &  716.5   &  5.5  $\pm$2.4 & 0.61 $\pm$ 0.22&2.5  $\pm$0.9  & 0.86$\pm$0.38& 2.33 $\pm$1.02& - & -   \\
A2631 		 &0.27 & 23h37m38.160s & +00d16m09.00s &  468.9   &  1234.6  &  6.9  $\pm$0.8 & 2.36 $\pm$ 0.23&9.9  $\pm$1.0  & 1.93$\pm$0.22& 7.05 $\pm$0.82& 0.28 $\pm$0.08 & 0.62 $\pm$0.19 \\    
\end{longtable}       
%}% End \longtab    
%\end{center}    
\end{landscape}

\end{document}